\documentclass[fleqn,usenatbib,usedcolumn]{mnras}

\usepackage[T1]{fontenc}
\usepackage{ae,aecompl}

\usepackage{multirow}
\usepackage{caption}
\usepackage{threeparttable}
\usepackage{dsfont}
\usepackage{xcolor}
\usepackage{arydshln}
\usepackage{commath}
\usepackage{graphicx}
\usepackage{bbold}

\usepackage{graphicx}  
\usepackage{amsmath}  
\usepackage{amssymb}  
\usepackage{csquotes}
\usepackage{anyfontsize}
\usepackage{physics}
\usepackage{newtxtext,newtxmath}
\usepackage{xspace}
\usepackage{booktabs}
\usepackage[capitalise]{cleveref}

\newcommand{\Mpc}{\,\hbox{Mpc}}
\newcommand{\Msol}{\,{\hbox{M}_\odot}}
\newcommand{\noop}[1]{}

\newcommand{\Msun}{\text{M}_\odot}

\newcommand{\eV}{\text{eV}}

\newcommand{\flamingo}{FLAMINGO}

\newcommand{\healpix}{\textsc{HEALPix}\xspace}

\newcommand{\diff}{\mathrm{d}}

\title[FLAMINGO: Thermal history of the Universe]{FLAMINGO: The thermal history of the Universe from tSZ effect cross-correlations and its dependencies on cosmology and baryon physics}

\author[Salcido et al.]{Jaime Salcido,$^{1}$
  Tianyi Yang,$^{1}$\thanks{E-mail: t.yang@ljmu.ac.uk}
  Ian G. McCarthy,$^{1}$\thanks{E-mail: i.g.mccarthy@ljmu.ac.uk},
  Emily E. Costello,$^{1}$
  Jonah T. Conley,$^{1}$\newauthor
  Willem Elbers,$^{2}$
  Carlos S. Frenk,$^{2}$
  Matthieu Schaller,$^{3,4}$
  Joop Schaye,$^{3}$
  Amol Upadhye,$^{5,6}$\newauthor
  Marcel~P.~van~Daalen,$^{3}$
  Bert~Vandenbroucke$^{3}$
  \\ \\
  $^{1}$ Astrophysics Research Institute, Liverpool John Moores University, 146 Brownlow Hill, Liverpool L3 5RF, UK\\
  $^{2}$Institute for Computational Cosmology, Department of Physics, University of Durham, South Road, Durham, DH1 3LE, UK\\
  $^{3}$Leiden Observatory, Leiden University, PO Box 9513, 2300 RA Leiden, the Netherlands\\
  $^{4}$Lorentz Institute for Theoretical Physics, Leiden University, PO box 9506, NL-2300 RA Leiden, the Netherlands\\
  $^{5}$~South-Western Institute for Astronomy Research, Yunnan University, Kunming, Yunnan 650500, People's Republic of China\\
  $^{6}$~Yunnan Key Laboratory of Survey Science, Yunnan University, Kunming, Yunnan 650500, People's Republic of China\\
}

\date{Accepted XXX. Received YYY; in original form ZZZ}

\pubyear{2026}

\begin{document}
\label{firstpage}
\pagerange{\pageref{firstpage}--\pageref{lastpage}}
\maketitle

\begin{abstract}
  The cross-correlation between tracers of large-scale structure, such as galaxies or quasars, and the thermal Sunyaev-Zel'dovich (tSZ) signal yields a measure of the bias-weighted mean electron pressure, $\langle b_\mathrm{h} P_\mathrm{e} \rangle$, where $b_\mathrm{h}$ is the halo bias and $P_\mathrm{e}$ is the electron pressure.  With a model for the bias, one can derive the thermal history, $\mathrm{d}y/\mathrm{d}z$, where $y$ is the Compton parameter and $z$ is redshift. We explore how these quantities depend on redshift, cosmology, and the physics of galaxy formation using the \flamingo\ suite of cosmological hydrodynamical simulations, which spans a range of cosmological parameters and baryonic feedback implementations in volumes of up to $(2.8\,\text{Gpc})^3$. We find that $\langle b_\mathrm{h} P_\mathrm{e} \rangle$ depends steeply on $S_8 \equiv \sigma_8\sqrt{\Omega_\mathrm{m}/0.3}$, with an effective scaling $\langle b_\mathrm{h} P_\mathrm{e} \rangle \propto S_8^{\epsilon(z)}$, where the exponent $\epsilon(z) \approx 3$ over the redshift range $0.1 \leq z \leq 1$. Compared with existing cross-correlation measurements using tracer samples from SDSS, BOSS, eBOSS, DES, and DESI cross-correlated with tSZ measurements from \textit{Planck}, we find that models with a low-$S_8$ cosmology and strong feedback are preferred, with a joint fit yielding $S_8 = 0.72^{+0.03}_{-0.03}$ and a normalised group-mass halo baryon fraction $f_\mathrm{b}(10^{13}\,\Msol,\,z{=}0.1)/(\Omega_\mathrm{b}/\Omega_\mathrm{m}) = 0.10^{+0.09}_{-0.05}$. Contrary to most probes of feedback which sample smaller scales (e.g., X-ray measurements), we show that feedback \textit{boosts} $\langle b_\mathrm{h} P_\mathrm{e} \rangle$, thus providing a novel test of feedback models.  Overall, our results show the thermal history provides a route to jointly constrain cosmological parameters and test models of galaxy formation.
\end{abstract}




\section{Introduction}\label{sec:intro}

The thermal energy content of the Universe encodes valuable information about the formation and evolution of cosmic structure. As baryons collapse into dark matter haloes, gravitational compression, shock heating, and astrophysical feedback processes raise the energy density of the gas, building up the cosmic reservoir of ionised plasma \citep[e.g.][]{cen_1999,fukugita_2004}. The redshift evolution of this thermal energy budget reflects an interplay between cosmological structure growth and baryonic feedback mechanisms such as from active galactic nuclei (AGN) and supernovae \citep[e.g.][]{benson_what_2003,bower_breaking_2006}. Cosmological parameters that regulate the growth of structure, such as the present-day matter density, $\Omega_\mathrm{m}$, the (linearly evolved) amplitude of the matter power spectrum filtered on $8\,h^{-1}\,\Mpc$ scales, $\sigma_8$, and the neutrino mass, alter the abundance and clustering of haloes, thereby also modifying the distribution of hot gas. At the same time, feedback processes redistribute thermal energy between virialised haloes and the diffuse intergalactic medium (IGM), whereas radiative cooling can reduce the thermal energy of the gas.  Studying the redshift evolution of the Universe's thermal energy therefore provides a powerful avenue for jointly testing cosmological models and models of galaxy formation \citep[e.g.][]{Chiang2020mzc,Chiang2021,Chen2024}.

The thermal Sunyaev–Zel'dovich (tSZ) effect \citep{Sunyaev1972} offers an ideal probe of this thermal energy reservoir. The tSZ effect arises from inverse Compton scattering of cosmic microwave background (CMB) photons off hot electrons, inducing a spectral distortion whose magnitude is characterised by the Compton parameter, $y$. This dimensionless quantity is proportional to the line-of-sight integral of the electron pressure:
\begin{equation}\label{eqn:y_params_definition}
  y=\frac{\sigma_{\rm T}}{m_{\mathrm{e}}c^2}\int P_{\mathrm{e}}\;\diff\ell \propto \int n_{\mathrm{e}} T_{\mathrm{e}}\;\diff\ell,
\end{equation}
where $\sigma_{\rm T}$ is the Thomson scattering cross-section, $m_{\mathrm{e}}c^2$ is the electron rest energy, $P_{\mathrm{e}} = n_{\mathrm{e}}k_{\rm B}T_{\mathrm{e}}$ is the electron pressure, $k_{\rm B}$ is the Boltzmann constant, and $n_\mathrm{e}$ and $T_{\mathrm{e}}$ are the electron number density and temperature, respectively. Because $y$ is sensitive to both halo gas and diffuse large-scale structures, the tSZ effect provides a more complete census of the thermal energy of ionised gas than probes such as X-ray emission, extending to lower-density environments and higher redshifts \citep[e.g.][]{SZ_summary_ref1,SZ_summary_ref2, SZ_summary_ref3}.

Observationally, the tSZ signal can be extracted either from intensity fluctuation maps produced by CMB anisotropy experiments such as \textit{Planck}, ACT, and SPT (e.g., \citealt{Planck_2015y,SPT_Planck_combined_22,ACT_y_24,SPT_y_new26}), or from absolute CMB spectrometers that constrain the sky-averaged $y$ monopole through measurements of spectral distortions of the primary CMB [e.g. with COBE-FIRAS \citep{COBE_FIRAS,new_FIRAS_measure}, or future experiments such as BISOU \citep{BISOU}, TMS \citep{TMS}, COSMO \citep{COSMO}, and FOSSIL \citep{fossil_link}].

Cross-correlating tSZ intensity maps with galaxy (or quasar) surveys yields a measurement of the halo bias-weighted mean electron pressure, $\langle b_\mathrm{h} P_\mathrm{e} \rangle$, as a function of redshift \citep[e.g.][]{Vikram2017dpo,Chiang2020mzc,Maleubre2026}. A rapidly growing ensemble of such cross-correlation measurements now exist \citep{Vikram2017dpo,pandey2019,Koukoufilippas2020,Chiang2020mzc,Chen2023_tsz,Sanchez2022,La_Posta_26}.  \citet{Chen2024} have shown using the halo model formalism that $\langle b_\mathrm{h} P_\mathrm{e} \rangle$ depends on key cosmological parameters, reporting $\langle b_\mathrm{h} P_\mathrm{e} \rangle \propto (\sigma_8 \Omega_\mathrm{m}^{0.81} h^{0.67})^{3.14}$ at redshift $z=0$.  However, the sensitivity to cosmology also raises the question of the sensitivity to baryon physics (and its uncertainties), and the potential degeneracies between cosmological effects and baryon physics.  Unlike weak lensing (cosmic shear), where stronger feedback suppresses the small-scale matter power spectrum \citep[e.g.][]{van_daalen_effects_2011,van_daalen_2020,McCarthy_2023,Salcido2023,Salcido2025}, in the present study we will show that the effect of feedback on $\langle b_\mathrm{h} P_\mathrm{e} \rangle$ is qualitatively different: stronger AGN feedback \emph{boosts} the large-scale pressure signal by redistributing thermal energy from halo cores to larger scales. Understanding and disentangling these two effects --- cosmology and feedback --- will therefore be essential for exploiting the thermal history as a cosmological probe (or for testing models of baryonic feedback).

Beyond the bias-weighted pressure, it is also useful to consider the quantity $\mathrm{d}y/\mathrm{d}z$, to which we refer to as the thermal history.  At a given redshift, the bias-weighted pressure is proportional to $\mathrm{d}y/\mathrm{d}z$, which itself is proportional to the mean thermal energy density of the Universe.  Deriving the thermal history from the bias-weighted pressure therefore requires a model for the bias, i.e. for how pressure fluctuations relate to matter fluctuations. Previous analyses typically used the halo model formalism \citep[as developed in][]{by_model_ref1,by_model_ref2,by_model_ref3,by_model_ref4,by_model_ref5} for this purpose.  However, uncertainties associated with these bias corrections have not been fully elucidated in previous studies. We anticipate that the bias will also depend on cosmology and baryon physics and therefore any mismatch between assumed and actual bias will propagate into systematic uncertainties in the inferred thermal energy history.

In this paper, we address these challenges using the \flamingo\ suite of cosmological hydrodynamical simulations \citep{Schaye2023,Kugel2023}. \flamingo\ combines Gpc-scale volumes --- necessary for robust large-scale statistics --- with systematic variations in both cosmological parameters and baryonic feedback prescriptions, all calibrated against observed galaxy and cluster properties. We measure $\langle b_\mathrm{h} P_\mathrm{e} \rangle$ from three-dimensional power spectra across 18 simulation variants spanning different values of $S_8 \equiv \sigma_8\sqrt{\Omega_\mathrm{m}/0.3}$, neutrino mass, and feedback strength, and we also calculate the $y$-weighted bias and thermal history $\mathrm{d}y/\mathrm{d}z$.

The main objectives of the present study are: (i) to quantify the cosmological and baryonic dependencies of $\langle b_\mathrm{h} P_\mathrm{e} \rangle$ and construct a compact parameterisation suitable for use in inference pipelines; (ii) to characterise the cosmic thermal history, $\mathrm{d}y/\mathrm{d}z$, via the bias evolution, clarifying which physical processes and scales dominate the observable signal at different epochs; and (iii) to explore how these signals are related to the redshift evolution of tSZ--halo mass scaling relations.

The paper is structured as follows. In \cref{sec:sims} we describe the \flamingo\ simulations and our methodology for computing four complementary thermal diagnostics: $\langle b_\mathrm{h} P_\mathrm{e} \rangle$, Compton $y$ maps, $\mathrm{d}y/\mathrm{d}z$, and the $y$-weighted halo bias. In \cref{sec:results} we present our results, including cosmological and baryonic dependence, parametric fitting functions, MCMC inference, and the decomposition of the tSZ signal. We summarise and discuss the implications of our findings in \cref{sec:conclusions}.

\section{Simulations and methodology}
\label{sec:sims}

\subsection{\flamingo\ simulations}

We provide here a brief summary of the \flamingo\ simulations. A detailed description of the simulations is presented in \citet{Schaye2023}.

\flamingo\ is a suite of large-scale cosmological hydrodynamical simulations designed to study cosmology and large-scale structure (LSS) physics. The suite includes three different mass resolutions: a high-resolution run with baryon particle mass $m_{\rm gas} = 1.3\times10^{8}~\rm M_{\odot}$ (referred to as m8), an intermediate-resolution run with $m_{\rm gas} = 1.1\times10^{9} ~\rm M_{\odot}$ (m9), and a low-resolution run with $8.6\times10^{9}~\rm M_{\odot}$ (m10). The flagship runs are the $(1 ~\rm Gpc)^{3}$ high-resolution run and the $(2.8 ~\rm Gpc)^{3}$ intermediate-resolution run (denoted as L1$\_$m8 and L2p8$\_$m9 respectively). The latter follows $2.8 \times 10^{11}$ particles, making it the largest cosmological hydrodynamical simulation evolved to $z=0$ at the time it was run. Most runs adopt a Dark Energy Survey Year Three \citep[DES Y3;][]{Abbott2022} 3 $\times$ 2pt + All Ext. $\Lambda$CDM cosmology (see \cref{tab:cosmologies} for the list of parameter values). This cosmology assumes a spatially-flat universe and is based on a combination of constraints from three DES Y3 two-point correlation functions along with external data.

The simulations were performed using \textsc{Swift} \citep{SWIFT_ref}, a fully open-source coupled cosmology, gravity, hydrodynamics, and galaxy formation code. The hydrodynamic equations are solved using the smoothed particle hydrodynamics (SPH) method \citep[for a review, see][]{SPH_review_price}, in particular the SPHENIX flavour of SPH \citep{Borrow_SPHENIX}, which was designed specifically for simulations of galaxy formation. The initial conditions are obtained from a modified version of monofonIC \citep{IC_ref1,IC_ref2}, and neutrinos are implemented with the $\delta f$ method \citep{FLAMINGO_neutrino}.

Unresolved physical processes are treated with subgrid prescriptions. The simulations include radiative cooling and heating \citep{FMG_radiative_heating_cooling}, star formation and evolution \citep{FMG_SF_ref_08S,Wiersma_09b}, black hole growth \citep{Booth_schaye_09,Bahe_22}, feedback from young stars and supernovae \citep{FMG_SF_ref_22b,FMG_SF_ref_22a}, and AGN feedback \citep{Booth_schaye_09,Husko_jet}.

\begin{table*}
  \centering
  \begin{threeparttable}[b]
    \caption{Cosmological parameters used in the simulations, reproduced from \citet{Schaye2023}. The columns list the prefix used to indicate the cosmology in the simulation name; the
      dimensionless Hubble constant, $h$; the total matter density parameter, $\Omega_\text{m}$; the dark energy density parameter, $\Omega_\Lambda$; the
      baryonic matter density parameter, $\Omega_\text{b}$; the sum of the particle masses of the neutrino species, $\sum m_\nu$; the amplitude of
      the primordial matter power spectrum, $A_\text{s}$; the power-law index of the primordial matter power spectrum, $n_\text{s}$; the amplitude of the
      initial power spectrum parametrized as the r.m.s. mass density fluctuation in spheres of radius $8~h^{-1}\,\Mpc$ extrapolated to $z=0$ using linear
      theory, $\sigma_8$; the amplitude of the initial power spectrum parametrized as $S_8\equiv \sigma_8\sqrt{\Omega_\text{m}/0.3}$; the neutrino matter density
      parameter, $\Omega_\nu \cong \sum m_\nu/(93.14~h^2\,\eV)$. Note that the values of the Hubble and density parameters are given at $z=0$. The values of the parameters that are listed in the
    last three columns have been computed from the other parameters.}
    \label{tab:cosmologies}
    \begin{tabular}{lcccccccccc}
      \hline
      Prefix & $h$ & $\Omega_\text{m}$ & $\Omega_\Lambda$ & $\Omega_\text{b}$ & $\sum m_\nu$ & $A_\text{s}$ & $n_\text{s}$ & $\sigma_8$ & $S_8$ & $\Omega_\nu$\\
      \hline
      D3A            & 0.681 & 0.306 & 0.694 & 0.0486 & 0.06~eV & $2.099\times 10^{-9}$ & 0.967 & 0.807 & 0.815 & $1.39\times 10^{-3}$\\
      Planck          & 0.673 & 0.316 & 0.684 & 0.0494 & 0.06~eV & $2.101\times 10^{-9}$ & 0.966 & 0.812 & 0.833 & $1.42\times 10^{-3}$\\
      PlanckNu0p24Var & 0.662 & 0.328 & 0.672 & 0.0510 & 0.24~eV & $2.109\times 10^{-9}$ & 0.968 & 0.772 & 0.807 & $5.87\times 10^{-3}$\\
      PlanckNu0p24Fix & 0.673 & 0.316 & 0.684 & 0.0494 & 0.24~eV & $2.101\times 10^{-9}$ & 0.966 & 0.769 & 0.789 & $5.69\times 10^{-3}$\\
      LS8             & 0.682 & 0.305 & 0.695 & 0.0473 & 0.06~eV & $1.836 \times 10^{-9}$ & 0.965 & 0.760 & 0.766 & $1.39\times 10^{-3}$\\
      \hline
    \end{tabular}
  \end{threeparttable}
\end{table*}

\begin{table*}
  \centering
  \caption{\flamingo\ hydrodynamical simulations used in this work, adapted and extended from \citet{Schaye2023}. The first four lines list the simulations that use the fiducial galaxy formation model and assume the fiducial cosmology (D3A), but use different volumes and resolutions. The remaining lines list the model variations, which all use a 1~Gpc box and intermediate resolution. The columns list the simulation identifier (where m8, m9 and m10 indicate $\log_{10}$ of the mean initial baryonic particle mass and correspond to high, intermediate, and low resolution, respectively; absence of this part implies m9 resolution); the number of standard deviations by which the observed stellar masses are shifted before calibration, $\Delta m_\ast$; the number of standard deviations by which the observed cluster gas fractions are shifted before calibration, $\Delta f_\text{gas}$; the AGN feedback implementation (thermal or jets); the comoving box side length, $L$; the number of baryonic particles, $N_\text{b}$ (which equals the number of CDM particles, $N_\text{CDM})$; the number of neutrino particles, $N_\nu$; the initial mean baryonic particle mass, $m_\text{g}$; the mean CDM particle mass, $m_\text{CDM}$; the Plummer-equivalent comoving gravitational softening length, $\epsilon_\text{com}$; the maximum proper gravitational softening length, $\epsilon_\text{prop}$; and the assumed cosmology which is specified in \cref{tab:cosmologies}.}
  \label{tab:simulations}
  \begin{tabular}{lrrllrrllrrl}
    \hline
    Identifier & $\Delta m_\ast$ & $\Delta f_\text{gas}$ & AGN & $L$ & $N_\text{b}$ & $N_\nu$ & $m_\text{g}$ & $m_\text{CDM}$ & $\epsilon_\text{com}$ & $\epsilon_\text{prop}$ & Cosmology \\
    & ($\sigma$) & ($\sigma$) && (cGpc) &&& ($\Msun$) & ($\Msun$)  & (ckpc) & (pkpc) \\
    \hline
    L1\_m8             & 0 & 0 & thermal & 1 & $3600^3$ & $2000^3$ & $1.34\times 10^8$ & $7.06\times 10^8$    & 11.2  & 2.85 & D3A \\
    L1\_m9             & 0 & 0 & thermal & 1 & $1800^3$ & $1000^3$ & $1.07\times 10^9$ & $5.65\times 10^9$    & 22.3  & 5.70 & D3A \\
    L1\_m10              & 0 & 0 & thermal & 1 & $900^3$ & $500^3$ & $8.56\times 10^9$ & $4.52\times 10^{10}$ & 44.6 & 11.40 & D3A \\
    L2p8\_m9           & 0 & 0 & thermal &2.8& $5040^3$ & $2800^3$ & $1.07\times 10^9$ & $5.65\times 10^9$    & 22.3  & 5.70 & D3A \\
    fgas$+2\sigma$  & 0 & $+2$ & thermal & 1 & $1800^3$ & $1000^3$ & $1.07\times 10^9$ & $5.65\times 10^9$ & 22.3  & 5.70 & D3A \\
    fgas$-2\sigma$  & 0 & $-2$ & thermal & 1 & $1800^3$ & $1000^3$ & $1.07\times 10^9$ & $5.65\times 10^9$    & 22.3  & 5.70 & D3A \\
    fgas$-4\sigma$  & 0 & $-4$ & thermal & 1 & $1800^3$ & $1000^3$ & $1.07\times 10^9$ & $5.65\times 10^9$    & 22.3  & 5.70 & D3A \\
    fgas$-8\sigma$  & 0 & $-8$ & thermal & 1 & $1800^3$ & $1000^3$ & $1.07\times 10^9$ & $5.65\times 10^9$    & 22.3  & 5.70 & D3A \\
    M*$-\sigma$ & $-1$ & 0 & thermal & 1 & $1800^3$ & $1000^3$ & $1.07\times 10^9$ & $5.65\times 10^9$    & 22.3  & 5.70 & D3A \\
    M*$-\sigma$\_fgas$-4\sigma$     & $-1$ & $-4$ & thermal & 1 & $1800^3$ & $1000^3$ & $1.07\times 10^9$ & $5.65\times 10^9$    & 22.3  & 5.70 & D3A \\
    Jet             & 0 & 0 & jets & 1 & $1800^3$ & $1000^3$ & $1.07\times 10^9$ & $5.65\times 10^9$    & 22.3  & 5.70 & D3A \\
    Jet\_fgas$-4\sigma$  & 0 & $-4$ & jets & 1 & $1800^3$ & $1000^3$ & $1.07\times 10^9$ & $5.65\times 10^9$    & 22.3  & 5.70 & D3A \\
    NoCooling  & -- & -- & -- & 1 & $1800^3$ & $1000^3$ & $1.07\times 10^9$ & $5.65\times 10^9$    & 22.3  & 5.70 & D3A \\
    Planck          & 0 & 0 & thermal & 1 & $1800^3$ & $1000^3$  & $1.07\times 10^9$ & $5.72\times 10^9$    & 22.3  & 5.70 & Planck\\
    PlanckNu0p24Var & 0 & 0 & thermal & 1 & $1800^3$ & $1000^3$ & $1.06\times 10^9$ & $5.67\times 10^9$    & 22.3  & 5.70 & PlanckNu0p24Var \\
    PlanckNu0p24Fix & 0 & 0 & thermal & 1 & $1800^3$ & $1000^3$ & $1.07\times 10^9$ & $5.62\times 10^9$    & 22.3  & 5.70 & PlanckNu0p24Fix \\
    LS8             & 0 & 0 & thermal & 1 & $1800^3$ & $1000^3$ & $1.07\times 10^9$ & $5.65\times 10^9$    & 22.3  & 5.70 & LS8 \\
    LS8$-$fgas$-8\sigma$             & 0 & 0 & thermal & 1 & $1800^3$ & $1000^3$ & $1.07\times 10^9$ & $5.65\times 10^9$    & 22.3  & 5.70 & LS8 \\
    \hline
  \end{tabular}
\end{table*}

Four subgrid parameters—two related to stellar feedback, one to black hole growth, and one to AGN feedback—are calibrated to match the observed present-day galaxy stellar mass function (GSMF) and the low-redshift gas mass fraction within $r_{500\mathrm{c}}$ for galaxy groups and clusters. Here, $r_{500\mathrm{c}}$ is the radius within which the mean matter overdensity is 500 times the critical density of the Universe. Machine learning-based emulators were used in the calibration \citep{Kugel2023}. The emulators were used not only to calibrate the fiducial model to observations, but also to produce the model variations discussed below.

To explore the feedback dependence of the thermal history, we make use of feedback model variants from a series of $(1 ~\rm Gpc)^{3}$ runs that vary the strength of stellar and/or AGN feedback. These runs share the same particle resolution as the fiducial model, but apply shifts to the observed galaxy stellar masses (for $M_{\ast}$) or cluster gas fractions (for $f_{\rm gas}$) during calibration. In the M*$-\sigma$ model, the observed stellar mass function was shifted by $0.14$ dex to lower stellar mass. The gas fraction variants (denoted as $f_{\rm gas}\pm N\sigma$) were calibrated to the observed gas mass fractions shifted by +2, -2, -4, -8 times the measurement error (i.e. the dispersion between different observational measurements), respectively. We note that in all model variations the machine-learning emulators recalibrate all four subgrid parameters simultaneously to match the shifted observables; therefore, the fgas and M* labels refer to the calibration target that was varied, not to a single feedback channel. In addition, we consider two models where AGN feedback is implemented using a kinetic jet-based model, rather than through isotropic thermal energy injection. The fiducial jet model is calibrated against the same set of observational data as the fiducial thermal AGN feedback model, but we also consider the stronger Jet\_fgas$-4\sigma$ model.  The majority of the feedback variation runs were carried out in the fiducial D3A cosmology (see \cref{tab:simulations}).

To study the cosmological dependence of the pressure statistics, we also include several cosmology variants from the $(1 ~\rm Gpc)^{3}$ runs (see \cref{tab:cosmologies} for the list of parameter values). Three of the alternative cosmologies we consider are variations on \citet{Planck2020cosmopars}: their best-fitting $\Lambda \rm CDM$ model with the minimum allowed neutrino mass, $\Sigma m_{\nu}c^{2}$= 0.06 eV (`Planck'); a model with a high neutrino mass, $\Sigma m_{\nu}c^{2}$ = 0.24 eV, (allowed at 95 per cent confidence by \textit{Planck}; \citealt{Planck2020cosmopars}) in which the other cosmological parameters take their corresponding best-fitting values from the \textit{Planck} MCMC chains (`PlanckNu0p24Var'); and a model with the same high neutrino mass, $\Sigma m_{\nu}c^{2}$ = 0.24 eV, that keeps all other parameters fixed to the values of model Planck, except for $\Omega_{\rm CDM}$ which was reduced in order to keep $\Omega_{\rm m}$ constant (`PlanckNu0p24Fix'). Note that for the latter model we fix the primordial power spectrum amplitude, $A_{\rm s}$, rather than $S_{8}$. All models with $\Sigma m_{\nu}c^{2}$ = 0.24 eV use three massive neutrino species of 0.08 eV each. Finally, we include a `lensing cosmology' from \citet[][denoted LS8]{Amon2023}. This model has a lower amplitude of the power spectrum, $S_{8}=0.766$, compared with 0.815 and 0.833 for the fiducial and Planck runs respectively, and provides a better match to some previous cosmic shear measurements, including from DES Y3 \citep{Abbott2022} and KiDS-1000 \citep{Heymans2021}. All cosmology-variation runs are evolved using the same calibrated feedback model as the fiducial run, with the exception of the LS8$-$fgas$-8\sigma$ run described below, which combines the LS8 cosmology with stronger feedback.

Additionally, we include a comparison of two new runs introduced in \citet{Ian_kSZ_feedback_FLAMINGO}, both in 1 Gpc boxes at intermediate resolution. The first is a run denoted `NoCooling', which sets the net radiative cooling + radiative heating rate to zero for gas where the net rate would have been negative (i.e., net cooling). Consequently, there is no radiative cooling and also no star formation or feedback present in this simulation. While obviously unrealistic, comparisons to this run are helpful for quantifying the impact of feedback and cooling. The second new run, denoted `LS8$-$fgas$-8\sigma$', is a strong feedback model in the LS8 lensing cosmology. This run provides an opportunity to explore the degeneracy between feedback and cosmology, via comparison to the fgas-8$\sigma$ run in the fiducial D3A cosmology.

The \flamingo\ suite also includes halo lightcone catalogues \citep{Helly2026}, in which haloes are identified with HBT-HERONS \citep{han_hbt+:_2018,ForouharMoreno2025} and their properties are computed with SOAP \citep{McGibbon2025_SOAP}. Black hole particles serve as tracers of subhalo positions on the lightcone, with halo properties drawn from the nearest snapshot in redshift. The \flamingo\ simulation data products are available through the FLAMINGO database\footnote{\href{https://flamingo.strw.leidenuniv.nl/data.html}{https://flamingo.strw.leidenuniv.nl/data.html}} \citep{Helly2026}.

\subsection{Methodology}

We analyse the thermal pressure signature in the \flamingo\ simulations using two complementary approaches. First, we compute the halo bias-weighted mean electron pressure $\langle b_\mathrm{h} P_\mathrm{e} \rangle$ from three-dimensional power spectra. Second, we analyse Compton $y$ maps produced from lightcones to quantify the integrated tSZ signal of all gas particles along the line of sight.

\subsubsection{Bias-weighted pressure and $\mathrm{d}y/\mathrm{d}z$ from 3D power spectra}

To measure $\langle b_\mathrm{h} P_\mathrm{e} \rangle$, we follow the methodology of \citet{Chiang2020mzc} and \citet{Young2021} using the 3D matter--electron pressure cross-spectrum and the matter power (auto) spectrum for each of the \flamingo\ simulations.  The description of how these power spectra are computed and corrected for shot noise is given in \citet{Schaye2023}.  Note that the 3D power spectra are output on the fly with a high redshift cadence, particularly at low redshifts where non-linear and baryonic effects are most evident.   Specifically, the simulations adopted an output frequency of $\Delta z = 0.05$ between $z=0$ and $z=3$ (60 outputs), $\Delta z = 0.25$ between $z=3$ and $z=12$ (36 outputs), $\Delta z = 0.5$ between $z=12$ and $z=20$ (16 outputs), and $\Delta z = 1$ between $z=20$ and $z=30$ (10 outputs).  We restrict our analysis to $z \le 3$ in the present study.

The bias-weighted mean electron pressure is defined through the large-scale limit of the ratio:
\begin{equation}
  \langle b_\mathrm{h} P_\mathrm{e} \rangle(z) = \lim_{k \to 0} \frac{P_{\mathrm{m} P_\mathrm{e}}(k,z)}{P_{\mathrm{mm}}(k,z)} \, .
  \label{eq:bhPe}
\end{equation}
\noindent where $P_{\mathrm{m} P_\mathrm{e}}(k,z)$ and $P_{\mathrm{mm}}(k,z)$ are the matter--electron pressure and matter--matter power spectra, respectively, and $b_\mathrm{h}(M,z)$ is the linear bias of dark matter haloes. Note that although no halo catalogs are required to compute the ratio of power spectra on the right hand side of \cref{eq:bhPe}, we still refer to the bias on the left hand side as a `halo' bias.  This is both for consistency with previous studies that modelled the tSZ cross-correlations using the halo model and because, in practice, massive haloes dominate the pressure field (as we demonstrate below).

$\langle b_\mathrm{h} P_\mathrm{e} \rangle$ is the fundamental observable in tSZ--galaxy (or quasar) cross-correlation studies \citep{Vikram2017dpo,Chiang2020mzc,Chen2024}. Physically, the electron pressure is a biased tracer of the matter field because it originates primarily from hot gas in massive, collapsed structures \citep{Refregier2000}. On large scales, the pressure fluctuations are well described by the linear bias expansion:
\begin{equation}
  \delta P_\mathrm{e}(\mathbf{k},z) \simeq \bar{P}_\mathrm{e}(z) b_y(z) \delta_{\mathrm{m}}(\mathbf{k},z),
  \label{eq:pressure_bias}
\end{equation}
where $b_y(z)$ is the $y$-weighted linear bias parameter and $\bar{P}_\mathrm{e}(z)$ the mean electron pressure, so that $\langle b_\mathrm{h} P_\mathrm{e} \rangle(z) = b_y(z) \bar{P}_\mathrm{e}(z)$. The cross-power spectrum between electron pressure and matter fluctuations then takes the form:
\begin{align}
  \langle \delta_{\mathrm{m}}(\mathbf{k}, z) P_\mathrm{e}(\mathbf{k}', z) \rangle &= (2\pi)^3 \delta_{\rm D}^{(3)}(\mathbf{k} + \mathbf{k}') P_{\mathrm{m} P_\mathrm{e}}(k, z), \label{eq:cross_power_general}\\
  P_{\mathrm{m} P_\mathrm{e}}(k, z) &\simeq \langle b_\mathrm{h} P_\mathrm{e} \rangle(z) P_{\mathrm{mm}}(k, z) \, , \label{eq:cross_power_bias}
\end{align}
making the ratio $P_{\mathrm{m} P_\mathrm{e}}/P_{\mathrm{mm}}$ scale-independent on large scales where linear bias applies.

\begin{figure}
  \centering
  \includegraphics[width=0.48\textwidth]{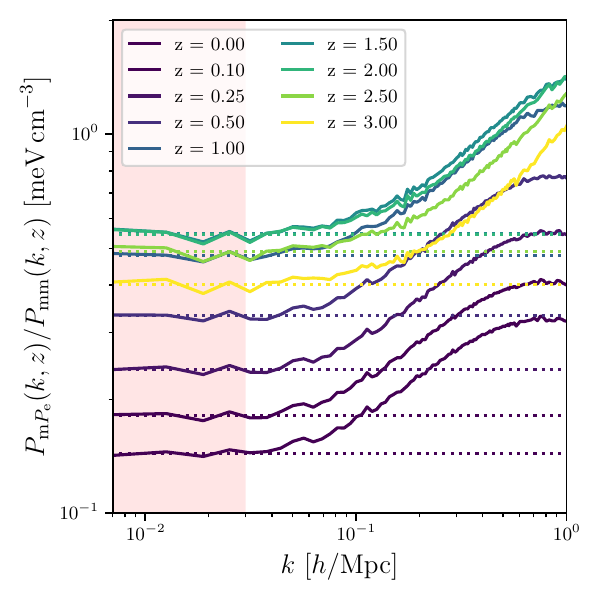}
  \caption{Scale dependence of the bias-weighted mean electron pressure in the fiducial L1\_m9 \flamingo\ simulation. The ratio $P_{\mathrm{m} P_\mathrm{e}}(k,z)/P_{\mathrm{mm}}(k,z)$ between the matter--electron pressure cross-power spectrum and matter auto-power spectrum is shown as a function of wavenumber $k$ for different redshifts. The horizontal dotted lines indicate the mean ratio averaged over $k < 0.03 \,h\,\mathrm{Mpc}^{-1}$ at each redshift, whilst the light shaded region highlights this averaging range. The scale-independent behaviour on large scales ($k < 0.03 \,h\,\mathrm{Mpc}^{-1}$) justifies the linear bias approximation and enables robust extraction of $\langle b_\mathrm{h} P_\mathrm{e} \rangle$ via \cref{eq:bhPe}. At smaller scales ($k > 0.1 \,h\,\mathrm{Mpc}^{-1}$), non-linear effects and one-halo contributions cause deviations from the large-scale limit, emphasising the importance of the scale cut for reliable bias-weighted pressure measurements.}
  \label{fig:cross}
\end{figure}

We average this ratio over $k < 0.03$ $h$/Mpc, restricting to the two-halo regime where the linear bias approximation holds. This scale cut matches observational analyses and avoids both finite box size effects and non-linear complications. The resulting scale-independent behaviour is confirmed at all redshifts in \cref{fig:cross}.

We compute $\langle b_\mathrm{h} P_\mathrm{e} \rangle$ for each simulation variant and compare with measurements from SDSS, BOSS, eBOSS, DES, and DESI surveys cross-correlated with \textit{Planck} Compton $y$ maps.

Using the bias-weighted pressure, we can compute the cosmic evolution of thermal energy, $\mathrm{d}y/\mathrm{d}z$, which follows directly from \cref{eqn:y_params_definition} and is related to $\langle b_\mathrm{h} P_\mathrm{e} \rangle$ (e.g., \citealt{Vikram2017dpo, Chiang2020mzc}) via:
\begin{equation}\label{eqn::b_y_P_e_dy_dz_relation}
  \langle b_\mathrm{h} P_\mathrm{e} \rangle = b_{y} \langle P_\mathrm{e} \rangle  = \frac{m_\mathrm{e} c^{2}(1+z)}{\sigma_{\rm T}}\frac{\mathrm{d}z}{\mathrm{d}\chi} \left( \frac{\mathrm{d}y}{\mathrm{d}z}b_{y} \right),
\end{equation}
where $\chi$ is the comoving distance to redshift $z$.

To isolate $\mathrm{d}y/\mathrm{d}z$, one then needs to correct for the pressure-weighted bias, $b_{y}$. Many previous studies have used the halo model formalism to compute the pressure-weighted bias via:
\begin{equation}\label{eqn::b_y_halo_model}
  b_{y}(z) = \frac{\int \diff M\frac{\diff n}{\diff M}M^{5/3+\alpha_{p}}b_\mathrm{h}(M,z)}{\int \diff M\frac{\diff n}{\diff M}M^{5/3+\alpha_{p}}},
\end{equation}
where $dn/dM$ is the halo mass function.  Note that the $M^{5/3}$ terms arise from the self-similar expectation for the dependence of pressure on halo mass and $\alpha_p$ is a factor introduced in \citet{Arnaud2010} to account for the mild deviation of the pressure profiles of observed galaxy groups and clusters from self-similar expectations.

Alternatively, one can use the simulations to directly evaluate $b_y$ as $\langle b_\mathrm{h} P_\mathrm{e} \rangle$ / $\bar{P}_\mathrm{e}$, where the volume-weighted mean electron pressure can be calculated from the particles as:
\begin{equation}\label{eqn:: averaged_P_e}
  \bar{P}_\mathrm{e} = \frac{k_{\rm B}}{V}\sum_{i=1}^{N} \Big(\frac{m_{\rm g,i}~n_{\rm e,i}~T_{\rm i}}{\rho_{\rm g,i}}\Big),
\end{equation}
where $m_{\rm g, i}$, $\rho_{\rm g,i}$, and $n_{\rm e,i}$ are the gas mass, gas density, and electron number density of the \textit{i}$^\mathrm{th}$ particle, and $V$ is the simulation volume.

Later we will compare the predictions of the halo model with the direct simulation calculation for $b_y$.  For the halo model prediction, we use the halo mass function prescription given in \citet{Tinker_HMF}, and the linear halo bias from \citet{Tinker_halo_bias}. We note that neither of these fitting functions provides a particularly accurate description of the \flamingo\ halo statistics \citep[see][]{Schaye2023,Maleubre2026}, but we adopt them here for consistency with previous work. We assume a fiducial D3A cosmology when computing $\textrm{d}n/\textrm{d}M$ and $b_\mathrm{h}(M,z)$. The parameter $\alpha_{p}$ characterises the small deviation of the electron pressure profiles from self-similarity as suggested by X-ray cluster observations, with a value of 0.12 \citep[see][]{Arnaud2010, Chiang2020mzc, Young2021}. In this study, however, we take $\alpha_{p}=0$ in order to compare our directly measured $b_{y}$ from the various \flamingo\ models with the self-similar halo model prediction.  In the integration, we adopt a mass range of $10^{11}/h~\textrm{M}_{\odot}<~M_{500\mathrm{c}}<5\times10^{15}/h ~\textrm{M}_{\odot}$, where $\rm M_{500\mathrm{c}}$ is the total mass enclosed within $r_{500\mathrm{c}}$.

To make a fair comparison between simulations, we use the simulation-derived $b_{y}$ from each model to correct the bias-weighted tomographic $y$. With the bias-corrected $\mathrm{d}y/\mathrm{d}z$, we can further estimate the sky-averaged $y$-monopole by integrating $\mathrm{d}y/\mathrm{d}z$ up to a given redshift. In this study, we integrate $\mathrm{d}y/\mathrm{d}z$ up to $z = 3.0$, and compare this value to the mean of the stacked $y$ lightcone map from gas particle snapshots. This upper integration limit is chosen because it corresponds to the maximum redshift available for the lightcone map outputs, but we find that the $y$ monopole is approximately converged by this upper redshift in any case.

\subsubsection{$\mathrm{d}y/\mathrm{d}z$ from lightcone-based maps}\label{ssec::y-statistics-from-maps}

A complementary, simulation-only approach for exploring the thermal history is via the full-sky \flamingo\ lightcone tSZ maps. Observationally, the Compton-$y$ signal is integrated along the full line of sight, so the tomographic $\mathrm{d}y/\mathrm{d}z$ must be inferred indirectly from cross-correlation measurements of $\langle b_\mathrm{h} P_\mathrm{e} \rangle$, which requires a model for the $y$-weighted bias $b_y$. In the simulations, however, one can construct redshift-resolved $y$ maps directly from the known positions and thermodynamic properties of every gas element. This provides a useful consistency check on the bias-corrected $\mathrm{d}y/\mathrm{d}z$ derived from $\langle b_\mathrm{h} P_\mathrm{e} \rangle$. Here we provide a brief summary of the lightcone data generation. A detailed description of the algorithm can be found in the appendix of \citet{Schaye2023}.

To produce the \healpix \citep{Healpix_ref} maps for each quantity, an observer's past lightcone is split into a set of concentric spherical shells, with a redshift interval of $\Delta z = 0.05$ from $z = 0$ to 3. Whenever a particle is found to have crossed the lightcone, we determine which shell it lies in at the time of crossing and accumulate the particle's contributions to the \healpix maps for that shell. In our study, we use full-sky particle lightcone \healpix maps per shell with $N_{\rm side}=4096$, extending out to $z = 3.0$, which is the maximum redshift available for the lightcone map outputs from the $(1 \rm Gpc)^{3}$ box fiducial run and its model variations. When a gas particle crosses the lightcone, we accumulate the following dimensionless quantity
\begin{equation}\label{eqn::FMG_y}
  y = \frac{\sigma_{\rm T}k_{\rm B}}{m_{\rm e}c^{2}}\frac{m_{\rm gas}n_{\rm e}T_{\rm e}}{\Omega_{\rm pix}d^{2}_{\rm A}\rho}
\end{equation}
to the map, where $\Omega_{\rm pix}$ is the solid angle of a \healpix pixel and $d_{\rm A}$ is the angular diameter distance to the observer. Since the gas particles have associated smoothing lengths, quantities derived from the gas are smoothed onto the \healpix maps. A detailed description of the smoothing scheme can be found in Appendix A of \citet{Schaye2023}.

From the full-sky lightcone map of each shell, the tomographic $\mathrm{d}y/\mathrm{d}z$ is simply given by the sky-averaged Compton $y$ divided by the shell width, $\Delta z=0.05$.

\begin{figure*}
  \centering
  \includegraphics[width=0.98\textwidth]{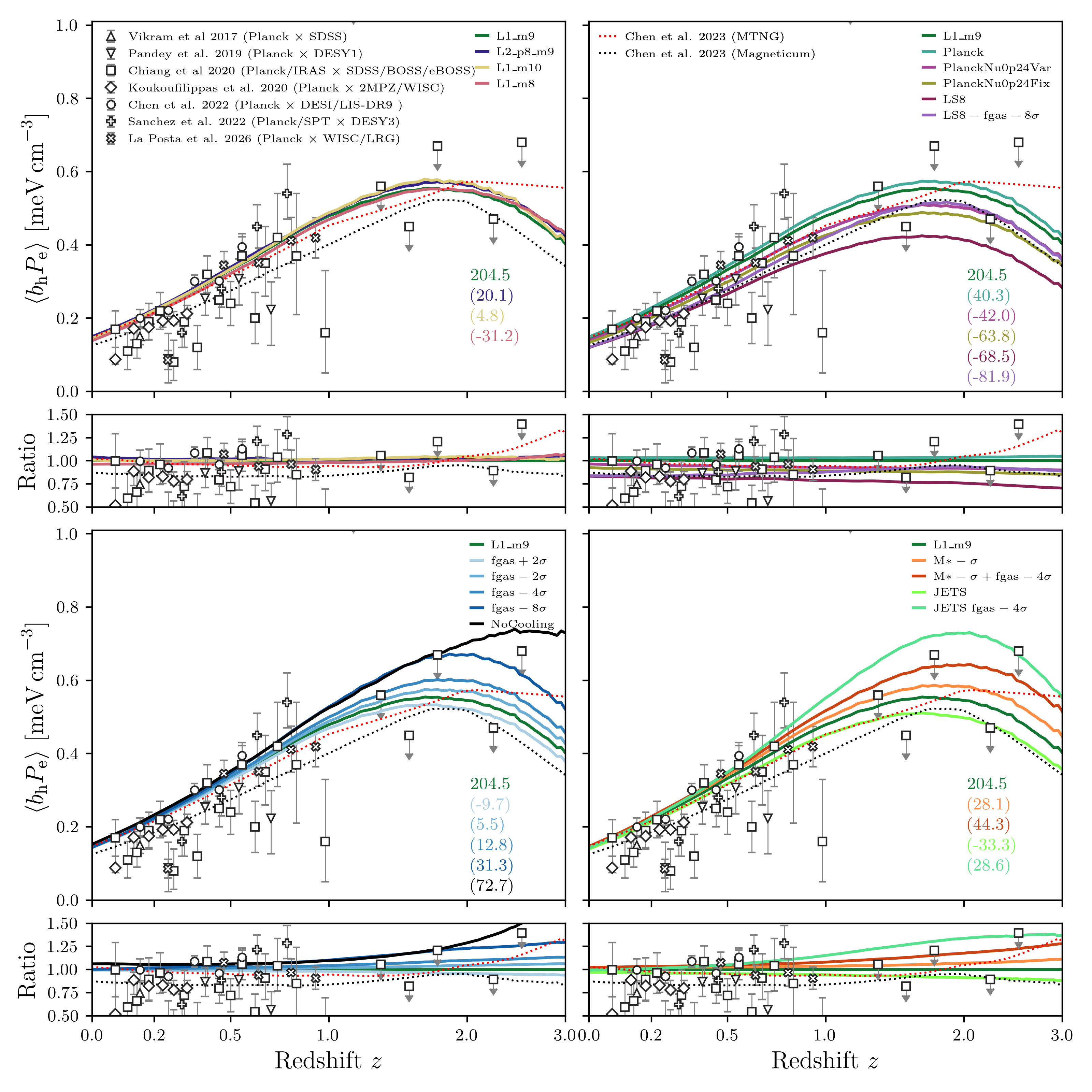}
  \caption{Bias-weighted mean electron pressure $\langle b_\mathrm{h} P_\mathrm{e} \rangle$ predicted by the \flamingo\ simulations compared with observational measurements. The symbols represent cross-correlation measurements between galaxy samples (SDSS groups, BOSS, eBOSS, DES, DESI) and thermal Sunyaev-Zel'dovich maps from \textit{Planck} and other CMB surveys, with error bars indicating observational uncertainties. The solid coloured lines show predictions from different \flamingo\ simulation variants. The numbers in the bottom right of each panel indicate $\chi^2$ values for the fiducial L1\_m9 simulation and $\Delta\chi^2$ differences (in parentheses) for other variants, obtained by directly comparing each simulation curve with the same 42 observational data points at $z<1$, treated as being independent; the corresponding reduced values are $\chi^2_{\rm r}=\chi^2/42$ and are summarised in \cref{tab:chi2_summary}. Negative $\Delta\chi^2$ values indicate improved agreement with observations. The bottom sub-panels show ratios relative to the fiducial L1\_m9 simulation, highlighting systematic differences. Dotted lines show results from the MillenniumTNG and Magneticum simulations for comparison. \textbf{Top left}: Resolution and box size dependence, demonstrating convergence between intermediate (L1\_m9) and high-resolution (L1\_m8) simulations. \textbf{Top right}: Cosmological parameter dependence, showing strong sensitivity to $\sigma_8$ and $\Omega_\mathrm{m}$ (or $S_8$), and behaving largely as an overall normalisation shift across redshift. The Planck cosmology predicts higher electron pressures (poorer fit), whilst the LS8 cosmology provides the best agreement with observations. \textbf{Bottom left}: Baryonic physics variations through feedback strength (via cluster gas fraction targets). In our calibrated model variations, lowering the target gas fractions (stronger feedback) increases $\langle b_\mathrm{h} P_\mathrm{e} \rangle$; the differences are small at $z\lesssim 1$ and become more pronounced at higher redshifts. \textbf{Bottom right}: Additional astrophysical variations including stellar mass function shifts and jet-mode AGN feedback.}
  \label{fig:Pe_data}
\end{figure*}

\section{Results}
\label{sec:results}

In this section we present our main findings. In \cref{ssec:bh_pe} we explore the bias-weighted pressure as a function of cosmology and baryonic physics and present a compact parameterisation for this dependence. In addition, we compare to existing observational measurements of this quantity.  In \cref{ssec:integrated_tsz} we examine the thermal history and $y$-weighted bias, including an examination of the accuracy of the standard halo model formalism for computing the $y$-weighted bias.  We also discuss implications for the $y$ monopole.  In \cref{ssec:scaling_relations} we examine the tSZ--halo mass scaling relation as function of redshift and aperture for measuring the tSZ signal, demonstrating that feedback boosts the integrated signal within large apertures at high redshift.

\subsection{Bias-weighted pressure}
\label{ssec:bh_pe}

We begin by exploring how the bias-weighted pressure, $\langle b_\mathrm{h} P_\mathrm{e} \rangle$, depends on redshift, cosmological parameters, baryonic physics, as well as box size and resolution.

The top left panel of \cref{fig:Pe_data} shows that $\langle b_\mathrm{h} P_\mathrm{e} \rangle$ is well converged between the intermediate (L1\_m9) and high (L1\_m8) resolution simulations, with differences below 5 per cent across the redshift range examined. The low-resolution run (L1\_m10) yields slightly higher values at $z > 1$, likely because it under-resolves lower-mass haloes that contribute significantly to the pressure field at high redshift \citep{Chiang2020mzc,Chen2024}. The large-box simulation (L2p8\_m9) is consistent with L1\_m9, confirming that finite box size effects are negligible for our analysis.

The top right panel of \cref{fig:Pe_data} shows the sensitivity of $\langle b_\mathrm{h} P_\mathrm{e} \rangle$ to cosmology variations. The Planck cosmology yields systematically higher $\langle b_\mathrm{h} P_\mathrm{e} \rangle$ than the fiducial D3A cosmology by $\approx$10-15 per cent, driven primarily by its higher $S_8$ (as we demonstrate below). Increasing the summed neutrino mass from 0.06~eV to 0.24~eV substantially reduces $\langle b_\mathrm{h} P_\mathrm{e} \rangle$ at all redshifts, as massive neutrinos suppress the abundance and clustering of massive haloes. The resulting suppression of $\approx 7$ per cent in PlanckNu0p24Fix relative to Planck is consistent with halo model predictions for the power suppression at the scales probed here. The LS8 cosmology ($S_8 = 0.766$) predicts the lowest values and provides the best agreement with observations.  Interestingly, the LS8 cosmology run with enhanced feedback (LS8$-$fgas$-8\sigma$) has a boosted $\langle b_\mathrm{h} P_\mathrm{e} \rangle$ with respect to the LS8 run with fiducial feedback, particularly at higher redshifts.  This is a point we will return to below.

To quantify the goodness of fit, we compute $\chi^2$ values for each variant by directly comparing the simulation predictions with the observational data from \citet{Vikram2017dpo,pandey2019,Chiang2020mzc,Koukoufilippas2020,Sanchez2022,Chen2023_tsz,La_Posta_26} for $z<1$, treating the measurements as independent. This is an approximation, since all of the cross-correlation analyses use \textit{Planck} Compton-$y$ maps, introducing shared systematic uncertainties; however, the tracer samples are drawn from distinct surveys and redshift ranges, so the statistical errors are largely independent. By restricting to $z<1$, we ensure that all data points used are detections with well-defined uncertainties (i.e., no upper limits). In total, this comparison uses 42 observational data points, and we also report the corresponding reduced values, $\chi^2_{\rm r}=\chi^2/42$. The absolute goodness of fit should nevertheless be interpreted cautiously. Even closely neighbouring redshift bins in the observational compilation can differ by more than would be expected from statistical fluctuations about a smoothly evolving thermal history, suggesting the presence of residual systematics and/or underestimated uncertainties. In addition, neglecting covariances between measurements that share the same \textit{Planck} Compton-$y$ maps will inflate the absolute $\chi^2$ values. No smoothly varying physical model should therefore be expected to yield a reduced $\chi^2$ close to unity for this compilation. We therefore use these comparisons primarily to assess the \textit{relative} performance of the models rather than the absolute goodness of fit. \Cref{tab:chi2_summary} summarises the $\chi^2$, $\Delta\chi^2$, and reduced $\chi^2$ values for all model variants shown in \cref{fig:Pe_data}.
The $\Delta\chi^2$ values in \cref{fig:Pe_data} indicate the change relative to the fiducial, with negative values corresponding to improved fits. The Planck cosmology systematically worsens the fit, while the LS8 cosmology improves the agreement.

\begin{table}
  \centering
  \caption{Summary of goodness-of-fit statistics for the \flamingo\ model variants shown in \cref{fig:Pe_data}, obtained by directly comparing each simulation variant with the same set of 42 observational measurements of $\langle b_\mathrm{h} P_\mathrm{e} \rangle$ at $z<1$ compiled from \citet{Vikram2017dpo,pandey2019,Chiang2020mzc,Koukoufilippas2020,Sanchez2022,Chen2023_tsz,La_Posta_26}, treated as independent. We also list the corresponding reduced values, $\chi^2_{\rm r}=\chi^2/42$. All $\Delta\chi^2$ values are defined relative to the fiducial L1\_m9 model. The LS8$-$fgas$-8\sigma$ model has the lowest $\chi^2_{\rm r}$, and is highlighted in bold.}
  \label{tab:chi2_summary}
  \begin{tabular}{lccc}
    \hline
    Model & $\chi^2$ & $\Delta\chi^2$ & $\chi^2_{\rm r}$ \\
    \hline
    L1\_m9 (fiducial) & 204.48 & 0.00 & 4.87 \\
    \hline
    \multicolumn{4}{c}{Resolution and box size variations} \\
    L2p8\_m9 & 224.55 & +20.07 & 5.35 \\
    L1\_m10 & 209.31 & +4.83 & 4.98 \\
    L1\_m8 & 173.24 & -31.24 & 4.12 \\
    \hline
    \multicolumn{4}{c}{Cosmology variations} \\
    Planck & 244.73 & +40.25 & 5.83 \\
    PlanckNu0p24Var & 162.50 & -41.98 & 3.87 \\
    PlanckNu0p24Fix & 140.63 & -63.85 & 3.35 \\
    LS8 & 136.03 & -68.45 & 3.24 \\
    LS8$-$fgas$-8\sigma$ & 122.61 & -81.87 & \textbf{2.92} \\
    \hline
    \multicolumn{4}{c}{Feedback variations} \\
    fgas$+2\sigma$ & 194.82 & -9.66 & 4.64 \\
    fgas$-2\sigma$ & 209.98 & +5.50 & 5.00 \\
    fgas$-4\sigma$ & 217.30 & +12.82 & 5.17 \\
    fgas$-8\sigma$ & 235.76 & +31.28 & 5.61 \\
    NoCooling & 277.20 & +72.72 & 6.60 \\
    \hline
    \multicolumn{4}{c}{Other astrophysical variations} \\
    M*$-\sigma$ & 232.53 & +28.05 & 5.54 \\
    M*$-\sigma$\_fgas$-4\sigma$ & 248.76 & +44.28 & 5.92 \\
    Jet & 171.20 & -33.28 & 4.08 \\
    Jet\_fgas$-4\sigma$ & 233.12 & +28.64 & 5.55 \\
    \hline
  \end{tabular}
\end{table}

\begin{figure*}
  \centering
  \includegraphics[width=0.98\textwidth]{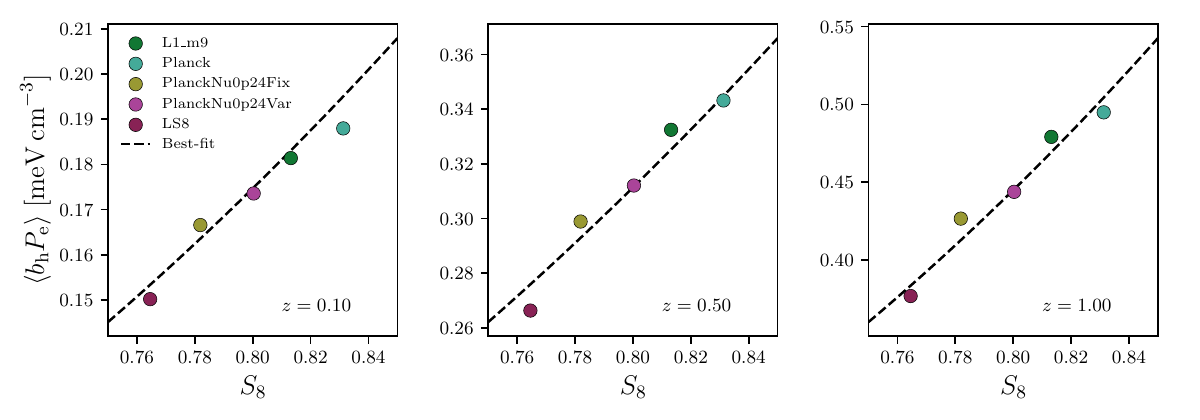}
  \caption{Cosmology scaling of the bias-weighted mean electron pressure with $S_8$ for a fixed feedback model. The three panels show $\langle b_\mathrm{h} P_\mathrm{e} \rangle$ measured in the \flamingo\ cosmology variants at $z=0.1$, 0.5, and 1.0, plotted against $S_8 \equiv \sigma_8\sqrt{\Omega_\mathrm{m}/0.3}$. The dashed curve in each panel shows the best-fitting form given by Eqs.~\eqref{eq:bhpe_s8_ansatz} and \eqref{eq:epsilon_s8_ansatz}. The relation is monotonic at each redshift, indicating that $\langle b_\mathrm{h} P_\mathrm{e} \rangle$ acts as a sensitive tracer of the amplitude of matter fluctuations.  All cosmology variations considered here adopt the fiducial \flamingo\ calibrated feedback model.}
  \label{fig:s8}
\end{figure*}

The strong cosmology dependence arises because $\langle b_\mathrm{h} P_\mathrm{e} \rangle$ depends on the abundance of massive haloes (scaling steeply with $\sigma_8$), their clustering strength, and their thermal energy content. For example, \citet{Chen2024} reported an effective scaling $\langle b_\mathrm{h} P_\mathrm{e} \rangle \propto (\sigma_8\,\Omega_\mathrm{m}^{0.81} h^{0.67})^{3.14}$ at $z=0$ from halo model calculations.

Motivated by the behaviour in \cref{fig:Pe_data}, where most cosmology changes appear largely as an overall normalisation shift, we summarise the cosmology dependence of the bias-weighted mean electron pressure with the fitting form
\begin{equation}
  \langle b_\mathrm{h} P_\mathrm{e} \rangle(z) = a\, S_8^{\epsilon(z)}\,(1+z)^\delta\, ,
  \label{eq:bhpe_s8_ansatz}
\end{equation}
where
\begin{equation}
  \epsilon(z) = \epsilon_0 + \epsilon_1 (1+z) + \epsilon_2 (1+z)^2\, ,
  \label{eq:epsilon_s8_ansatz}
\end{equation}
$a$ is a normalisation, $S_8\equiv \sigma_8\sqrt{\Omega_\mathrm{m}/0.3}$, and $\delta$ captures the remaining smooth redshift evolution. We fit this form to the \flamingo\ cosmology variants over $0.05 \leq z \leq 1$.
The quadratic form for $\epsilon(z)$ is the lowest-order polynomial that provides an accurate fit; simpler (constant or linear) forms leave systematic residuals. The best-fitting coefficients for both this cosmology-only model and the extended model including baryon fraction are summarised in \cref{tab:s8_fb_fit_params}. For the cosmology-only model, the effective exponent $\epsilon(z)$ remains close to $\approx 3$ over $0.1 \leq z \leq 1$ (ranging from 2.7 to 3.3), consistent with the scaling reported by \citet{Chen2024} from halo model calculations at $z=0$.

\Cref{fig:s8} highlights a tight scaling between the large-scale pressure observable and $S_8$. On the two-halo scales used here, $\langle b_\mathrm{h} P_\mathrm{e} \rangle$ is dominated by the abundance and clustering bias of the massive haloes that host most of the hot gas. Changing cosmology primarily rescales the mass function and halo bias, producing an approximately one-parameter family of predictions that tracks $S_8$.  As we discuss immediately below, the separation between cosmology and baryonic effects is relatively clean at low redshifts, where a measurement of $\langle b_\mathrm{h} P_\mathrm{e} \rangle$ can constrain $S_8$ almost independently of uncertainties in feedback modelling.

Regarding the dependence on baryon physics, the bottom left panel of \cref{fig:Pe_data} shows the impact of varying feedback strength through systematic shifts in the target cluster gas fractions. In these calibrated variations, lowering the target gas fractions (through stronger feedback) produces \emph{higher} $\langle b_\mathrm{h} P_\mathrm{e} \rangle$. The ratio panels reveal that baryonic differences are relatively small at $z \lesssim 1$ but increase with redshift as feedback becomes more active and $\langle b_\mathrm{h} P_\mathrm{e} \rangle$ becomes sensitive to lower mass haloes (see \citealt{Chiang2020mzc,Chen2024}) which are more susceptible to feedback effects.

The trend of increasing $\langle b_\mathrm{h} P_\mathrm{e} \rangle$ with decreasing gas fractions may appear counter-intuitive at first, but it is important to consider the scales involved.  The feedback in these fgas variation simulations was adjusted to reproduce shifted versions of the observed gas mass fractions measured within $r_{500\mathrm{c}}$, as inferred from X-ray observations.  From a clustering point of view, the simulations are therefore calibrated on the gas fractions deep in the one-halo regime.  And, indeed, reducing the gas fractions within $r_{500\mathrm{c}}$ does also lead to a reduction in the integrated pressure (or integrated Compton $y$) within that same scale (see, e.g., \citealt{LeBrun2015}).  However, the heated and ejected mass is redistributed on larger scales, which leads to an enhancement in $\langle b_\mathrm{h} P_\mathrm{e} \rangle$ relative to a model with weaker feedback (noting that the bias-weighted pressure is measured in the two-halo limit).  In \cref{ssec:scaling_relations} we explicitly demonstrate this by examining $Y$--$M$ scaling relations in different apertures.

\begin{figure*}
  \centering
  \includegraphics[width=0.98\textwidth]{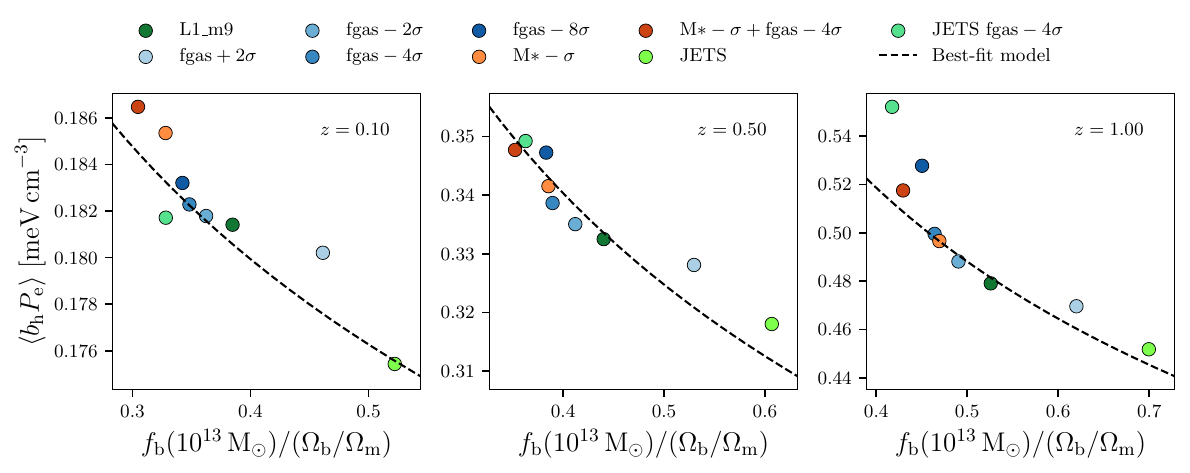}
  \caption{Bias-weighted mean electron pressure $\langle b_\mathrm{h} P_\mathrm{e} \rangle$ as a function of the group-mass halo baryon fraction $f_\mathrm{b}(10^{13}\,\Msol)$, normalised by the cosmic mean $\Omega_\mathrm{b}/\Omega_\mathrm{m}$. The three panels show results at $z=0.1$, 0.5, and 1.0. Points indicate the \flamingo\ feedback and subgrid-physics variants at fixed cosmology. The dashed curve in each panel shows the best-fitting form of \cref{eq:bhpe_s8_fb_ansatz}, evaluated at the fiducial $S_8$, and thus isolates the baryonic dependence of the parameterisation. The approximately monotonic trend demonstrates that $f_\mathrm{b}$ captures the leading-order baryonic dependence of the large-scale pressure observable, motivating its inclusion as a baryonic parameter in the joint parameterisation of \cref{eq:bhpe_s8_fb_ansatz}.
  }
  \label{fig:s8_fb_all_13_500}
\end{figure*}

Relative to the observational data, the stronger feedback variants tend to mildly worsen agreement. This is qualitatively different from cosmic shear, where stronger feedback is often favoured because it suppresses small-scale matter power and can improve agreement with the observed shear signal (e.g., \citealt{AE2022,McCarthy_2023}). However, it is important to highlight that the feedback variations in the bottom left panel of \cref{fig:Pe_data} were all run in a fixed cosmology, the fiducial D3A cosmology, which has a relatively high amplitude with respect to the data even at very low redshifts which are insensitive to feedback variations.  As increasing the feedback boosts the signal at higher redshifts, this only exacerbates the overall tension with the observational measurements.  However, if one instead considers the LS8 cosmology, as in the top right panel, we see that stronger feedback (i.e., LS8$-$fgas$-8\sigma$) is actually preferred over the fiducial feedback model in the standard LS8 run.  Indeed, the LS8$-$fgas$-8\sigma$ run provides the best fit to the observational measurements of any of the \flamingo\ simulations, which is consistent with the findings of recent studies examining the tSZ power spectrum \citep{Ian_kSZ_feedback_FLAMINGO,SPT2026} and the tSZ--cosmic shear cross-spectrum (\citealt{Ian_kSZ_feedback_FLAMINGO}; Yamamoto et al., in prep), as well as kSZ effect stacking analyses of massive galaxies (e.g., \citealt{Schaan2021,Bigwood2024,Hadzhiyska2025,Ian_kSZ_feedback_FLAMINGO,Siegel2026}).

Interestingly, the NoCooling model in the D3A cosmology has the highest $\langle b_\mathrm{h} P_\mathrm{e} \rangle$, particularly at high redshift.  This is not due to feedback redistributing mass from the one-halo to the two-halo regime, as this process is not present in that simulation.  On the contrary, the elevated pressure is driven by the absence of radiative cooling and star formation, which converts $\approx15$--$20$ per cent of the baryons to stars in the other simulations.  Thus,  cooling, star formation and feedback all influence the bias-weighted pressure and its evolution with redshift.

The bottom right panel of \cref{fig:Pe_data} shows additional variations (stellar mass function shifts and jet-mode AGN feedback). The jet model yields a modestly improved $\Delta\chi^2$ relative to the fiducial case, suggesting that the mode of energy coupling (thermal versus kinetic) can alter the redshift evolution of the pressure field in ways not captured by varying feedback strength alone.

Compared to predictions from other simulations, the fiducial \flamingo\ curve at low redshift is broadly consistent with the MillenniumTNG \citep[MTNG,][]{MTNG_ref1,MTNG_ref2} prediction, but somewhat higher than that from the Magneticum simulation suite \citep{mag_ref,Dolag2025}. This is expected, as MTNG adopts a \textit{Planck} 2015 cosmology ($S_8=0.828$, \citealt{Planck15_cosmo}), whereas Magneticum uses a WMAP7 cosmology with $S_8=0.770$ \citep{WMAP_cosmo}. At higher redshift, the predictions from FLAMINGO, MTNG, and Magneticum diverge more significantly. This is likely driven by differences in the implementation of baryon physics.

To better visualise the dependence of the bias-weighted pressure on baryon physics (at fixed cosmology) in FLAMINGO, \cref{fig:s8_fb_all_13_500} shows $\langle b_\mathrm{h} P_\mathrm{e} \rangle$ at three redshifts against the baryon fraction of group-mass haloes, $f_\mathrm{b}(10^{13}\,\Msol, z)$, normalised by the cosmic mean. The relation is roughly monotonic: lower baryon retention corresponds to increased large-scale pressure amplitudes across the entire redshift range. The NoCooling model is excluded from this figure because it does not follow the same monotonic relation. In the NoCooling case, the elevated $\langle b_\mathrm{h} P_\mathrm{e} \rangle$ is not driven by feedback-induced gas redistribution but by the absence of radiative cooling and star formation, which leaves a larger fraction of baryons in the hot gas phase. Consequently, its baryon fraction is close to the cosmic mean while its $\langle b_\mathrm{h} P_\mathrm{e} \rangle$ is among the highest, placing it well off the trend defined by the other models. Furthermore, the remaining models are all calibrated against the same observed stellar mass function, ensuring comparable stellar mass fractions, whereas the NoCooling model lacks star formation entirely; a parameterisation in terms of $f_\mathrm{b}$ alone cannot capture both regimes.

Motivated by this behaviour, we extend the cosmology fit by introducing a baryonic parameter based on the baryon fraction retained by group-mass haloes. We define
\begin{equation}
  f_\mathrm{b}(M_{500\mathrm{c}}, z) \equiv \frac{M_{\mathrm{b}}(<r_{500\mathrm{c}})}{M_{500\mathrm{c}}}\, ,
\end{equation}
and use a pivot mass of $M_{500\mathrm{c}}=10^{13}\,\Msol$. We tested a range of pivot masses, aperture definitions (e.g.\ $r_{200\mathrm{c}}$, $r_{100\mathrm{c}}$), and also considered the gas fraction $f_{\mathrm{gas}}$ (excluding the stellar contribution) in place of the total baryon fraction; the combination of $f_\mathrm{b}$ measured within $r_{500\mathrm{c}}$ at a pivot mass of $10^{13}\,\Msol$ yields the tightest description of the feedback dependence across the \flamingo\ variants, while also corresponding to a mass regime where feedback is expected to be particularly efficient at regulating gas content. Baryon masses within $r_{500\mathrm{c}}$ of the spherical overdensity (SO)-defined haloes are summed for the calculation. We then consider the phenomenological model
\begin{equation}
  \langle b_\mathrm{h} P_\mathrm{e} \rangle(z) = a\,\left[\frac{f_\mathrm{b}(10^{13}\,\Msol, z)}{\Omega_\mathrm{b}/\Omega_\mathrm{m}}\right]^{\gamma(z)} S_8^{\epsilon(z)}\,(1+z)^\delta\, ,
  \label{eq:bhpe_s8_fb_ansatz}
\end{equation}
with
\begin{equation}\label{eq:gamma_feedback_ansatz}
  \gamma(z) = \gamma_0 + \gamma_1 (1+z) + \gamma_2 (1+z)^2\, ,
\end{equation}
and $\epsilon(z)$ as defined in \cref{eq:epsilon_s8_ansatz}. Here the bracketed term measures the baryon fraction of $10^{13}\,\Msol$ haloes relative to the cosmic mean. The exponent $\gamma(z)$ encodes how changes in baryon retention propagate into the pressure observable, while $\epsilon(z)$ retains the cosmological dependence. This form treats the detailed complexity of feedback as being encapsulated, to leading order, by the group-scale baryon fraction. Furthermore, \cref{eq:bhpe_s8_fb_ansatz} implicitly assumes that the effects of cosmology and feedback are separable, i.e.\ multiplicative \citep[e.g.][]{mummery_separate_2017,pfeifer_2020,stafford_2020, Elbers2025}.

Fitting the \flamingo\ variants over $0.05 \leq z \leq 1$ yields the coefficients listed in \cref{tab:s8_fb_fit_params}, which summarises both the cosmology-only and extended cosmology-plus-baryon-fraction parameterisations. \Cref{fig:test_fit} in Appendix \ref{sec:validation} illustrates the comparison for a representative subset of nine variants chosen to span the full range of feedback strengths and cosmologies, including those with the largest residuals. Evaluated across all 14 simulation variants, the mean absolute relative error is $1.7 \pm 1.3$ per cent (median 1.4 per cent), with a worst-case deviation of $\approx 7$ per cent. The model residuals show a slight increase in scatter with redshift (i.e., they are mildly heteroscedastic): the mean error is $\approx 1.4$ per cent at low redshift ($z \leq 0.3$) and grows to $\approx 2.1$ per cent at $z \approx 0.7$--$1$, where the interplay between cosmology and feedback is strongest.

\begin{table*}
  \centering
  \caption{Posterior median coefficients and 68 per cent credible intervals for the two parameterisations of $\langle b_\mathrm{h} P_\mathrm{e} \rangle$ considered in this work, both calibrated over $0.05 \leq z \leq 1$. The first row corresponds to the cosmology-only model defined by \cref{eq:bhpe_s8_ansatz,eq:epsilon_s8_ansatz}. The second row corresponds to the extended model including the baryon-fraction dependence defined by \cref{eq:bhpe_s8_fb_ansatz,eq:gamma_feedback_ansatz}. Here $a$ is the overall normalisation, $\epsilon_i$ give the polynomial coefficients of the redshift-dependent $S_8$ scaling, $\gamma_i$ give the polynomial coefficients of the redshift-dependent baryon-fraction scaling, and $\delta$ captures the remaining smooth redshift evolution.}
  \label{tab:s8_fb_fit_params}
  \resizebox{\textwidth}{!}{%
    \begin{tabular}{lcccccccc}
      \hline
      Model & $a$ & $\epsilon_0$ & $\epsilon_1$ & $\epsilon_2$ & $\delta$ & $\gamma_0$ & $\gamma_1$ & $\gamma_2$ \\
      \hline
      Cosmology only & $0.282 \pm 0.002$ & $6.65 \pm 0.12$ & $-5.56 \pm 0.11$ & $1.937 \pm 0.025$ & $1.714 \pm 0.015$ & -- & -- & -- \\
      Cosmology + baryon fraction & $0.246 \pm 0.003$ & $3.67 \pm 0.45$ & $-1.80 \pm 0.64$ & $0.86 \pm 0.22$ & $1.783 \pm 0.039$ & $0.76 \pm 0.11$ & $-1.10 \pm 0.16$ & $0.30 \pm 0.06$ \\
      \hline
  \end{tabular}}
\end{table*}

\begin{figure}
  \centering
  \includegraphics[width=0.48\textwidth]{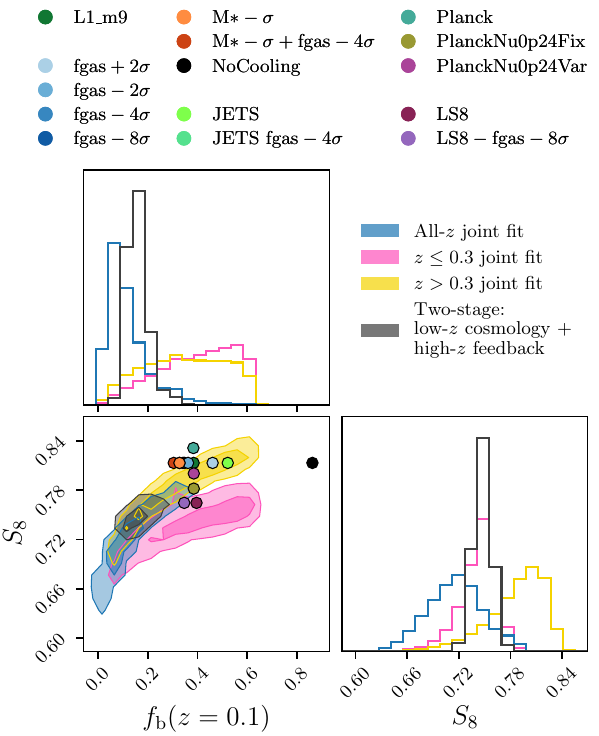}
  \caption{Comparison of posteriors from fits to the observed $\langle b_\mathrm{h} P_\mathrm{e} \rangle$ measurements shown in \cref{fig:Pe_data}. The first three analyses apply the full model, \cref{eq:bhpe_s8_fb_ansatz}, to different data subsets: all redshifts (blue), $z\leq0.3$ only (pink), and $z>0.3$ only (yellow). The two-stage analysis (grey) first constrains $S_8$ from the low-redshift data using the cosmology-only model, \cref{eq:bhpe_s8_ansatz}, where baryonic effects contribute less than 4 per cent variation, then uses the resulting $S_8$ posterior as a prior when fitting the full model to the high-redshift data ($z>0.3$). Contours show the 68 and 95 per cent credible regions projected onto $S_8$ and the baryon fraction $f_\mathrm{b}(10^{13}\,\Msol,\,z{=}0.1)/(\Omega_\mathrm{b}/\Omega_\mathrm{m})$. Coloured points indicate \flamingo\ simulation variants. The low-redshift data tightly constrain $S_8$ but not $f_\mathrm{b}$, while the high-redshift data are sensitive to both parameters but with a strong $f_\mathrm{b}$--$S_8$ degeneracy oriented in a different direction; combining both regimes breaks this degeneracy. All analyses consistently favour low $S_8$ and reduced baryon retention.}
  \label{fig:compare_posteriors}
\end{figure}

Having established the parameterisation, we demonstrate its use for joint inference by imposing a simple power-law ansatz for the redshift dependence of the group-scale baryon fraction within $r_{500\mathrm{c}}$:
\begin{equation}
  \frac{f_\mathrm{b}(M_{500\mathrm{c}}{=}10^{13}\,\Msol, z)}{\Omega_\mathrm{b}/\Omega_\mathrm{m}} = f_{\mathrm{b},0}(1+z)^\eta\, ,
  \label{eq:fb_mcmc_powerlaw}
\end{equation}
where $f_{\mathrm{b},0}\equiv f_\mathrm{b}(M_{500\mathrm{c}}{=}10^{13}\,\Msol, z{=}0)/(\Omega_\mathrm{b}/\Omega_\mathrm{m})$ is the baryon fraction ratio at $z=0$ and $\eta$ controls the redshift evolution. Substituting into \cref{eq:bhpe_s8_fb_ansatz} yields a three-parameter model $(f_{\mathrm{b},0}, \eta, S_8)$, which we fit to the observed $\langle b_\mathrm{h} P_\mathrm{e} \rangle$ data with a Gaussian likelihood, sampling the posterior via MCMC. The eight coefficients of the calibrated model (\cref{tab:s8_fb_fit_params}) are jointly sampled as nuisance parameters with a multivariate Gaussian prior centred on their posterior medians and covariance from the calibration fit, thereby marginalising over the model-parameter uncertainties. The resulting posterior medians (16th--84th percentile intervals) for the sampled parameters are $f_{\mathrm{b},0} = 0.09^{+0.09}_{-0.05}$, $\eta = 0.33^{+0.23}_{-0.22}$, and $S_8 = 0.72^{+0.03}_{-0.03}$. For comparison with the \flamingo\ simulations, \cref{fig:compare_posteriors} shows the posterior projected onto $S_8$ and the \emph{derived} baryon fraction at $z=0.1$, $f_\mathrm{b}(10^{13}\,\Msol, z{=}0.1)/(\Omega_\mathrm{b}/\Omega_\mathrm{m}) = f_{\mathrm{b},0}(1+0.1)^\eta$. Because $\eta$ is small, this is very close to $f_{\mathrm{b},0}$: $f_\mathrm{b}(10^{13}\,\Msol,\,z{=}0.1)/(\Omega_\mathrm{b}/\Omega_\mathrm{m}) = 0.10^{+0.09}_{-0.05}$. Both the low $S_8$ and the strongly reduced baryon retention in group-mass haloes are consistent with the direct goodness-of-fit comparison among \flamingo\ variants. We caution, however, that the derived best-fit values of $S_8$ and the group baryon fraction lie outside the ranges explored with the \flamingo\ simulations and therefore rely on an extrapolation of the model in \cref{eq:bhpe_s8_fb_ansatz}.  Nevertheless, these results are qualitatively consistent with our direct goodness-of-fit estimates for the simulations, which demonstrate that the LS8$-$fgas$-8\sigma$ run provides the best match to the existing observational measurements.

To illustrate how different redshift regimes contribute to these constraints, \cref{fig:compare_posteriors} also shows posteriors obtained by applying the full model separately to the low-redshift ($z\leq0.3$) and high-redshift ($z>0.3$) data; all four sets of constraints are summarised in \cref{tab:posterior_summary}. The low-redshift data tightly constrain $S_8 = 0.75^{+0.02}_{-0.02}$ but leave the baryon fraction essentially unconstrained ($f_\mathrm{b}(10^{13}\,\Msol,\,z{=}0.1)/(\Omega_\mathrm{b}/\Omega_\mathrm{m}) = 0.42^{+0.14}_{-0.19}$), consistent with the weak sensitivity of $\langle b_\mathrm{h} P_\mathrm{e} \rangle$ to feedback at low redshift discussed above. Conversely, the high-redshift data alone yield broader constraints on both parameters ($S_8 = 0.79^{+0.02}_{-0.04}$, $f_\mathrm{b}(10^{13}\,\Msol,\,z{=}0.1)/(\Omega_\mathrm{b}/\Omega_\mathrm{m}) = 0.35^{+0.18}_{-0.18}$) owing to the $f_\mathrm{b}$--$S_8$ degeneracy that strengthens with increasing redshift. Crucially, the degeneracy directions are markedly different: the low-redshift posterior is nearly horizontal in the $f_\mathrm{b}$--$S_8$ plane, reflecting a tight $S_8$ constraint that is largely independent of $f_\mathrm{b}$, whereas the high-redshift posterior is tilted, since increasing $S_8$ can be partially compensated by changes in the baryon fraction. The all-redshift joint fit therefore tightens the constraints on $f_\mathrm{b}$ by combining these complementary orientations, even though neither regime alone can simultaneously constrain both parameters well.

This complementarity motivates a two-stage analysis that directly exploits the clean separation between cosmology and baryonic effects at low redshift. At $z\leq0.3$, the total spread in $\langle b_\mathrm{h} P_\mathrm{e} \rangle$ across the full range of \flamingo\ feedback variants is less than 4 per cent, well below both the typical observational uncertainties ($\approx 10$--$30$ per cent) and the cosmology sensitivity ($\approx 20$ per cent between the Planck and LS8 cosmologies). The low-redshift data therefore constrain $S_8$ nearly independently of feedback uncertainties. In the first stage, we fit the cosmology-only model, \cref{eq:bhpe_s8_ansatz}, to the $z\leq0.3$ data, obtaining $S_8 = 0.75 \pm 0.01$, and in the second stage we use this posterior as a Gaussian prior when fitting the full model, \cref{eq:bhpe_s8_fb_ansatz}, to the complementary high-redshift data ($z>0.3$), where the baryon fraction becomes important. This yields $S_8 = 0.75^{+0.01}_{-0.01}$ and $f_\mathrm{b}(10^{13}\,\Msol,\,z{=}0.1)/(\Omega_\mathrm{b}/\Omega_\mathrm{m}) = 0.15^{+0.04}_{-0.03}$ (\cref{fig:compare_posteriors}). The two-stage $S_8$ is slightly higher and more tightly constrained than the joint fit ($S_8 = 0.72^{+0.03}_{-0.03}$), reflecting the additional broadening from the $f_\mathrm{b}$--$S_8$ degeneracy that enters when the full redshift baseline is used. This behaviour is consistent with the split-redshift fits described above: the $z\leq0.3$ data alone yield a comparably tight $S_8$ whether or not the baryon fraction is included in the model, validating the assumption that feedback uncertainties are negligible at low redshift. The two-stage baryon fraction is correspondingly higher and sits comfortably within the range spanned by the \flamingo\ simulation grid. Both analyses consistently favour a low-$S_8$ cosmology and reduced baryon retention in group-mass haloes; the agreement between the two independent approaches lends confidence to the robustness of these conclusions.

\begin{table}
  \centering
  \caption{Posterior median constraints (16th--84th percentile intervals) on $S_8$ and baryon fraction $f_\mathrm{b}(10^{13}\,\Msol,\,z{=}0.1)/(\Omega_\mathrm{b}/\Omega_\mathrm{m})$ from the four analyses shown in \cref{fig:compare_posteriors}. The all-redshift, $z\leq0.3$, and $z>0.3$ fits all use the full model, \cref{eq:bhpe_s8_fb_ansatz}. The two-stage analysis first constrains $S_8$ from the $z\leq0.3$ data using the cosmology-only model, \cref{eq:bhpe_s8_ansatz}, then fits the full model to the $z>0.3$ data with the resulting $S_8$ posterior as a prior.}
  \label{tab:posterior_summary}
  \renewcommand{\arraystretch}{1.4}
  \begin{tabular}{lcc}
    \hline
    Analysis & $S_8$ & $f_\mathrm{b}/(\Omega_\mathrm{b}/\Omega_\mathrm{m})$ \\
    \hline
    All-$z$ joint fit          & $0.72^{+0.03}_{-0.03}$ & $0.10^{+0.09}_{-0.05}$ \\
    $z\leq0.3$ joint fit       & $0.75^{+0.02}_{-0.02}$ & $0.42^{+0.14}_{-0.19}$ \\
    $z>0.3$ joint fit          & $0.79^{+0.02}_{-0.04}$ & $0.35^{+0.18}_{-0.18}$ \\
    Two-stage                  & $0.75^{+0.01}_{-0.01}$ & $0.15^{+0.04}_{-0.03}$ \\
    \hline
  \end{tabular}
  \renewcommand{\arraystretch}{1.0}
\end{table}

\begin{figure}[t]
  \centering
  \includegraphics[width=0.48\textwidth]{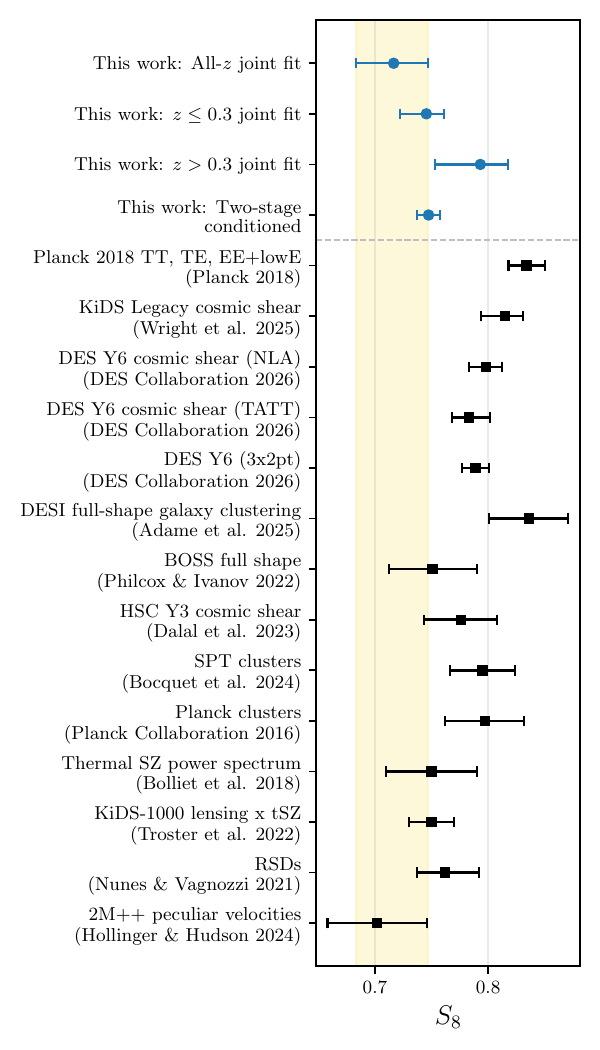}
  \caption{Comparison of $S_8$ constraints from this work with recent determinations from independent cosmological probes. Blue circles show posterior medians and 68 per cent credible intervals from our analyses (all-$z$ joint fit, $z\leq0.3$ fit, $z>0.3$ fit, and the two-stage conditioned analysis; see \cref{tab:posterior_summary}). Black squares show literature constraints from \textit{Planck} 2018 CMB, KiDS Legacy cosmic shear, DES Y6 cosmic shear (NLA and TATT), DES Y6 3$\times$2pt, DESI full-shape galaxy clustering, BOSS full shape, HSC Y3 cosmic shear, SPT and \textit{Planck} cluster counts, tSZ power spectrum, KiDS-1000 lensing$\,\times\,$tSZ, redshift-space distortions, and peculiar velocity measurements \citep{Planck2020cosmopars,Wright2025,DESY6,DESY6_3x2,Adame2025,Philcox2022,Dalal2023,Bocquet2024,Planck2016,Bolliet2018,Troster2022,Nunes2021,Hollinger2024}. The yellow shaded band indicates the 68 per cent credible interval from our all-$z$ joint fit. Our constraints consistently favour a lower $S_8$ than the primary CMB, but are compatible with several large-scale structure probes, including recent tSZ-based measurements.}
  \label{fig:s8_comparison}
\end{figure}

\Cref{fig:s8_comparison} places our inferred $S_8$ constraints in the context of recent determinations from independent cosmological probes. Our all-redshift joint fit lies below the primary-CMB result from Planck and toward the low-amplitude end of recent large-scale-structure constraints, while the two-stage result shifts upward and is correspondingly closer to several recent low-redshift measurements, including other tSZ-based statistics (e.g., \citealt{Bolliet2018,Troster2022}; see also \citealt{McCarthy_2023,Ian_kSZ_feedback_FLAMINGO,SPT2026}).

\begin{figure*}
  \centering
  \includegraphics[width=0.98\linewidth]{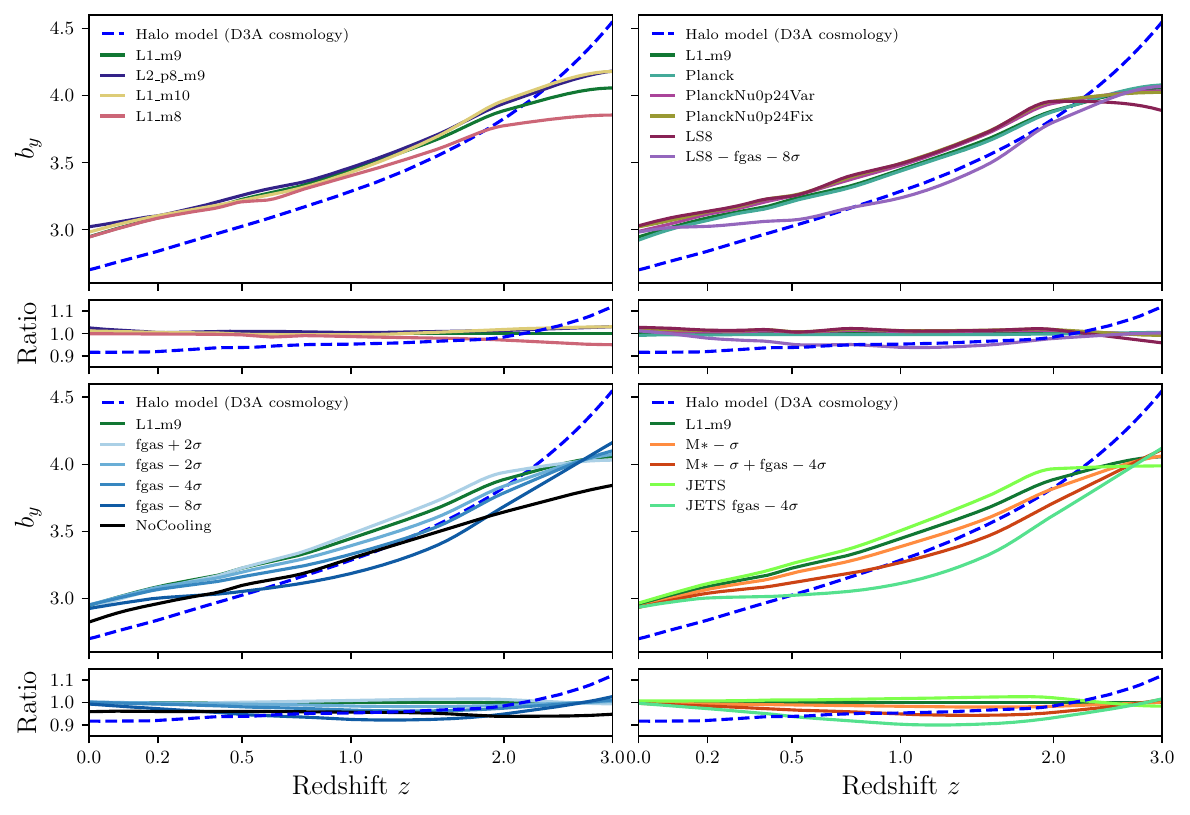}
  \caption{Evolution of the $y$-weighted halo bias $b_{y}$ as a function of redshift from different \flamingo\ model variations. This quantity reflects the clustering strength of haloes that dominate the thermal Sunyaev-Zel'dovich signal. The blue dashed curve shows the self-similar halo-model prediction given by \cref{eqn::b_y_halo_model}, assuming $\alpha_{p}=0.0$. Overall, the bias increases with redshift at $z\lesssim2$ but turns over and falls below the halo-model prediction at higher redshift. \textbf{Top left}: Resolution and box size dependence. \textbf{Top right}: Cosmology variations, with the LS8 cosmology exhibiting slightly higher bias values. \textbf{Bottom left}: Baryonic physics variations through feedback strength (via cluster gas fraction variations). Stronger feedback models show lower $b_{y}$ values. \textbf{Bottom right}: Other astrophysical variations including stellar mass function shifts and jet-mode AGN feedback.}
  \label{by_comparison_plots}
\end{figure*}

\subsection{Thermal history, \texorpdfstring{$\mathrm{d}y/\mathrm{d}z$}{dy/dz}, and the \texorpdfstring{$y$}{y} monopole}
\label{ssec:integrated_tsz}

We now turn our attention to the thermal history.  As discussed in \cref{sec:sims}, to estimate the (unbiased) thermal history $\mathrm{d}y/\mathrm{d}z$, one requires a model (or measurement) of the $y$-weighted bias, $b_y$; see \cref{eqn::b_y_P_e_dy_dz_relation}.  Here we compute $b_y$ as $\langle b_\mathrm{h} P_\mathrm{e} \rangle(z)$ / $\bar{P}_\mathrm{e}(z)$ and we compare the results with the predictions of the self-similar halo model.

\Cref{by_comparison_plots} shows the $y$-weighted halo bias, $b_{y}$, computed from different \flamingo\ model variations. Overall, $b_{y}$ increases with redshift for $z\lesssim2.0$ across all models, which implies that the thermal energy becomes increasingly associated with more highly biased haloes at higher redshift. This behaviour is broadly consistent with the predictions of the self-similar halo model, computed from \cref{eqn::b_y_halo_model}.

Focusing on individual models, the top right panel shows the cosmology dependence of $b_{y}$. Compared to the fiducial and \textit{Planck}-like cosmologies, the LS8 model, as well as the cosmology with increased summed neutrino mass, produce slightly higher bias amplitude. In these cosmologies, structure growth is suppressed, leading to a reduced abundance of massive clusters and weaker overall halo clustering. As a result, the haloes that do host significant thermal energy become more biased tracers of the underlying pressure field. In contrast, the fiducial and Planck cosmologies predict a higher abundance of massive haloes and stronger clustering, which lowers the effective bias of the thermal energy field.

Regarding baryonic physics (at fixed cosmology), stronger feedback (lower target gas fractions) redistributes more thermal energy from massive haloes into the surrounding medium, thereby reducing the bias of the thermal energy field. In comparison, the M*$-\sigma$ model and the kinetic jet model behave more similarly to the fiducial model, indicating a weaker impact on the large-scale bias. This trend is consistent with what is observed in the $\langle b_\mathrm{h} P_\mathrm{e} \rangle$ measurements.

It is noticeable that the simulated $b_{y}(z)$ curves exhibit a turnover and fall below the halo model prediction at $z\gtrsim2.5$. A similar finding was reported by \citet{Young2021} using the \textit{Magneticum} simulation. One possible explanation is that, at high redshift, feedback more efficiently redistributes hot gas into the diffuse IGM. In this regime, a substantial fraction of the thermal energy resides in the field rather than in bound haloes, whereas the halo model implicitly assumes that the pressure field is composed entirely of haloes.

\begin{figure*}
  \centering
  \includegraphics[width=0.98\linewidth]{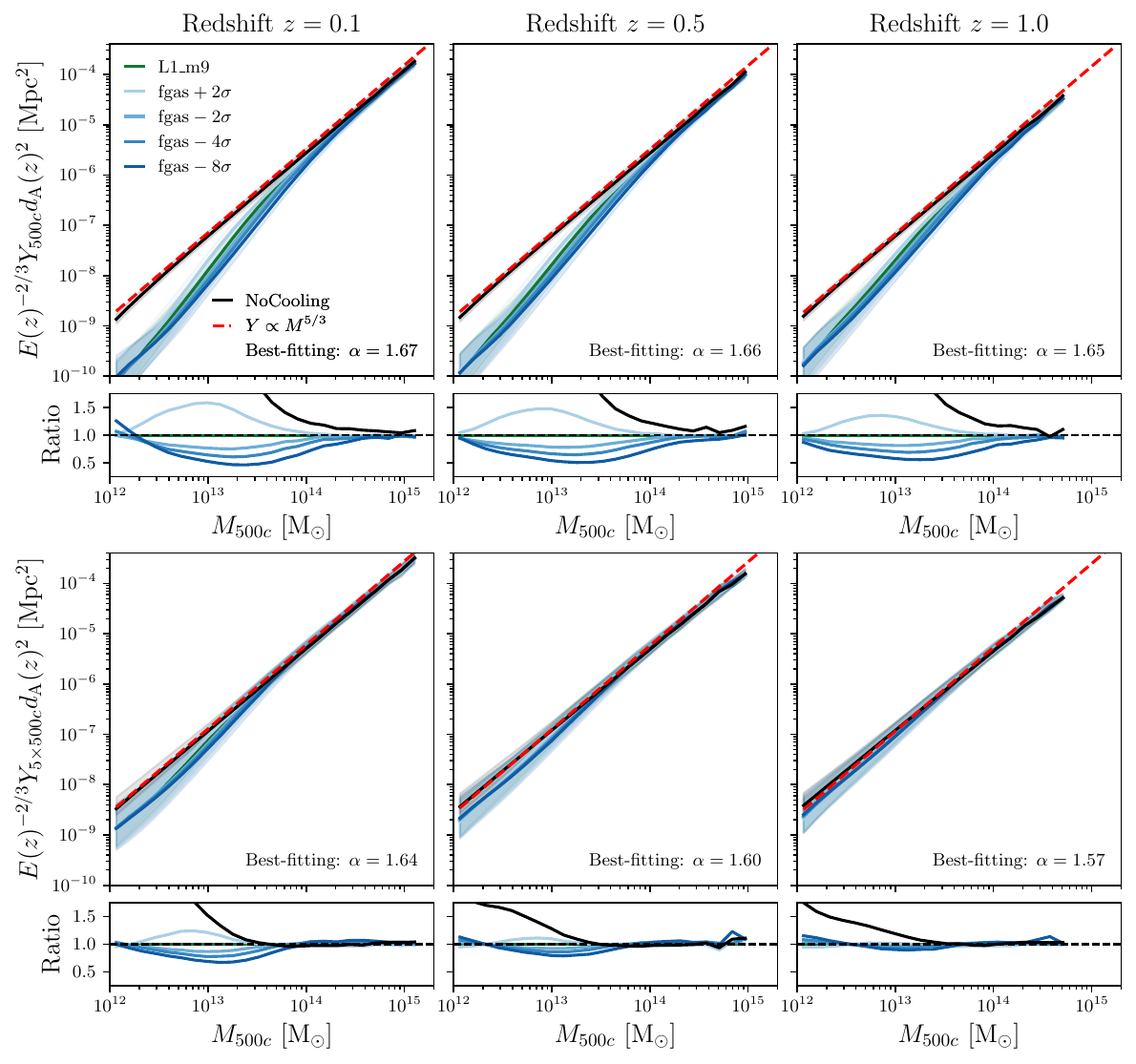}
  \caption{The median $Y$--$M$ relation for baryonic physics variations through feedback strength (via gas fraction variations), computed within $r_{500\mathrm{c}}$ (\textbf{top row}) and $5r_{500\mathrm{c}}$ (\textbf{bottom row}) for spherical overdensity (SO)-defined haloes. \textbf{Left}: $z=0.1$. \textbf{Middle}: $z=0.5$. \textbf{Right}: $z=1.0$. The fiducial model is shown in green, the gas fraction variations in blue, the NoCooling run in black, and the self-similar relation as a red dashed line. The shaded region shows the 16th--84th percentile range around the median. The bottom sub-panels display the ratio between the $Y$--$M$ relation for each gas fraction variation and that of the fiducial run. The suppression of electron pressure is more pronounced in the one-halo regime than in the two-halo regime. For comparison, predictions from the NoCooling and self-similar models are overplotted. The best-fitting mass-scaling power-law index for the NoCooling model is indicated in each legend panel. It is clear that the self-similar assumption in the NoCooling model breaks down when contributions from the two-halo regime are included.}
  \label{Y_M_fgas_test_0_1}
\end{figure*}

\begin{figure*}
  \centering
  \includegraphics[width=0.98\linewidth]{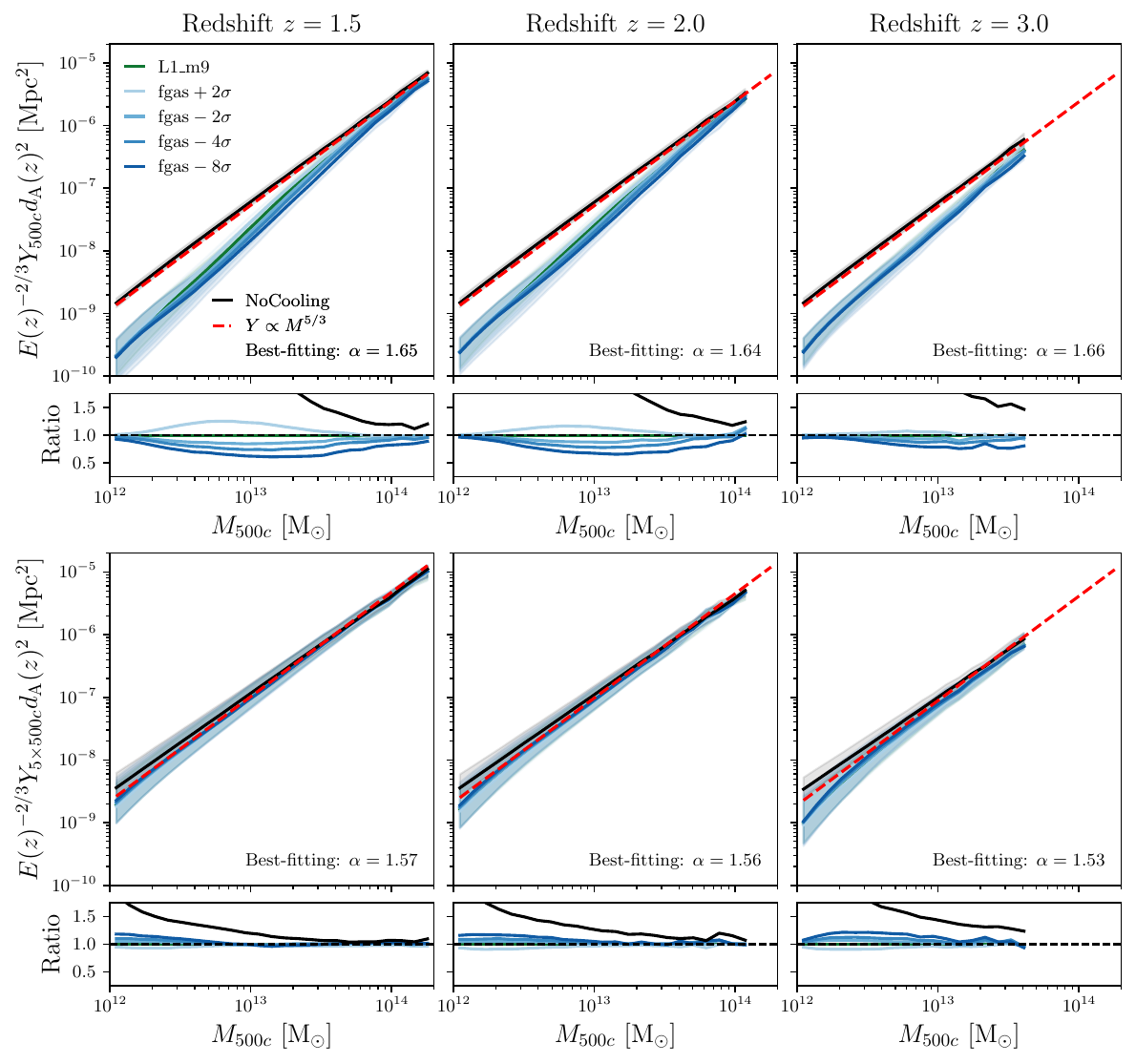}
  \caption{As \cref{Y_M_fgas_test_0_1}, but at higher redshift. \textbf{Left}: $z=1.5$. \textbf{Middle}: $z=2.0$. \textbf{Right}: $z=3.0$. A clear enhancement in electron pressure is seen for models with stronger feedback at higher redshift.}
  \label{Y_M_fgas_test_1.5_3}
\end{figure*}

However, this explanation cannot account for the NoCooling model, which lacks feedback capable of ejecting gas from haloes, yet still deviates from the halo model prediction at high redshift. A more likely driving factor is the breakdown of the self-similar assumption adopted for the pressure profile in the halo model. Indeed, as we will show later (in \cref{Y_M_fgas_test_0_1} and \cref{Y_M_fgas_test_1.5_3}), the scaling relation between the integrated tSZ effect and halo mass ($Y$--$M$) does increasingly deviate from the self-similar prediction of $M^{5/3}$ with increasing redshift, particularly when the tSZ effect is integrated within large apertures (e.g., $5 r_{500\mathrm{c}}$ vs $r_{500\mathrm{c}}$).  We find that the halo model computation of $b_y$ is very sensitive to the assumed slope of the $Y$--$M$ relation and the deviations from self-similarity we observe in the simulation scaling relations are sufficient to explain the deviation in $b_y(z)$ from the predictions of the halo model.

\begin{figure*}
  \centering
  \includegraphics[width=0.98\linewidth]{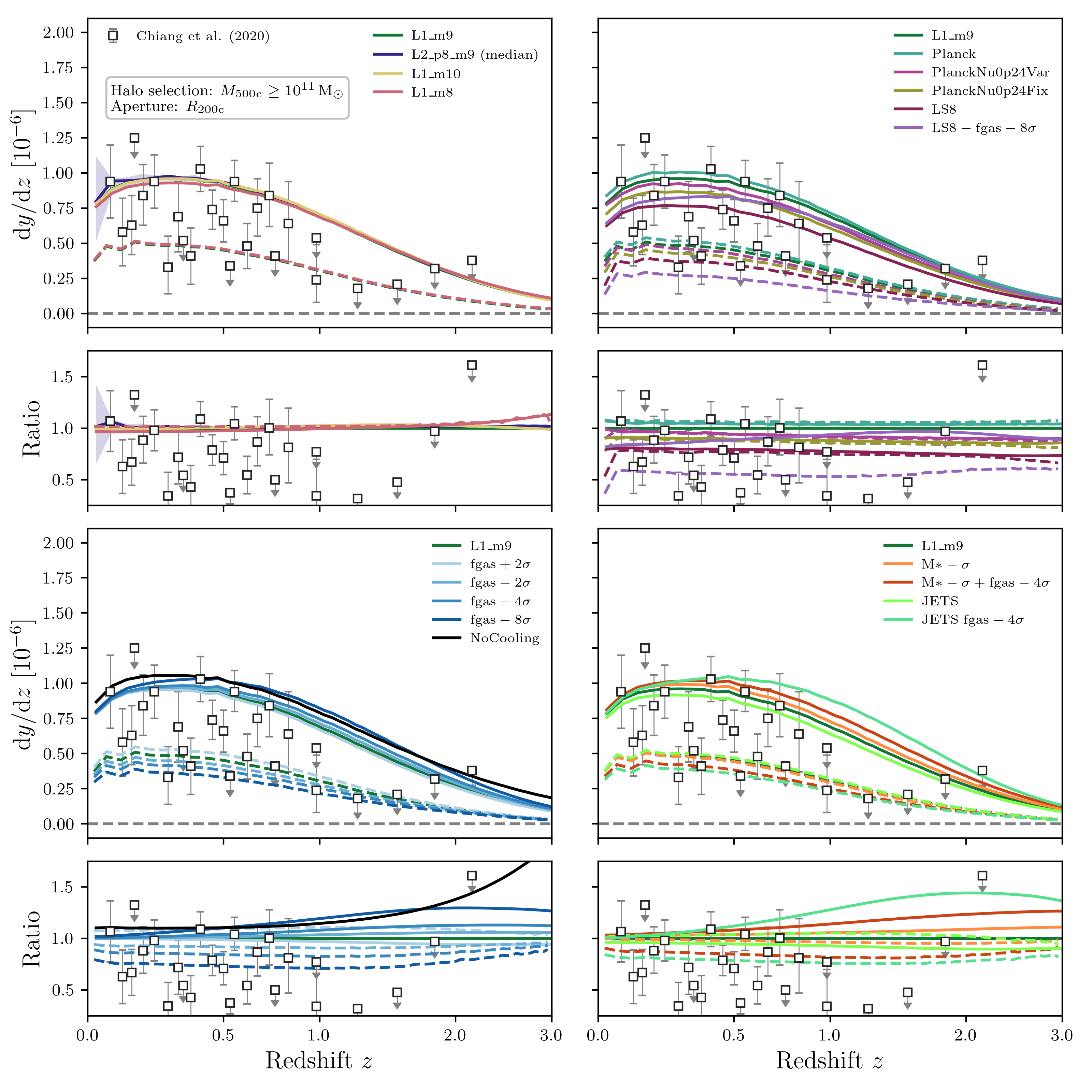}
  \caption{Redshift distribution of the thermal Sunyaev-Zel'dovich signal $\mathrm{d}y/\mathrm{d}z$ from the \flamingo\ simulations. This quantity represents the contribution of different redshift shells to the sky-averaged $y$ monopole, tracing the cosmic thermal history through structure formation. The solid curves show the direct prediction from the full-sky Compton-$y$ maps, while the dashed curves show the corresponding halo-based reconstruction for haloes with $M_{500\mathrm{c}}\geq 10^{11}\,\Msol$ using the $r_{200\mathrm{c}}$ aperture. \textbf{Top left}: convergence tests. \textbf{Top right}: cosmology variations. \textbf{Bottom left}: Feedback variations. \textbf{Bottom right}: other astrophysical variations including stellar mass function shifts and jet-mode AGN feedback. Data points from \citet{Chiang2020mzc}, derived from cross-correlation measurements using a halo-model estimate of $b_y$, are shown for reference; as these are model-dependent inferences rather than direct measurements, they should not be used to rule out specific models. The dashed curves lie systematically below the solid ones, indicating a non-negligible contribution from gas outside $r_{200\mathrm{c}}$; this is quantified further in \cref{sec:dydz_radius}. The peak contribution occurs around $z \approx 0.5$, and the signal declines rapidly at higher redshifts.}
  \label{dydz_comparison_plots}
\end{figure*}

\begin{figure*}
  \centering
  \includegraphics[width=0.98\linewidth]{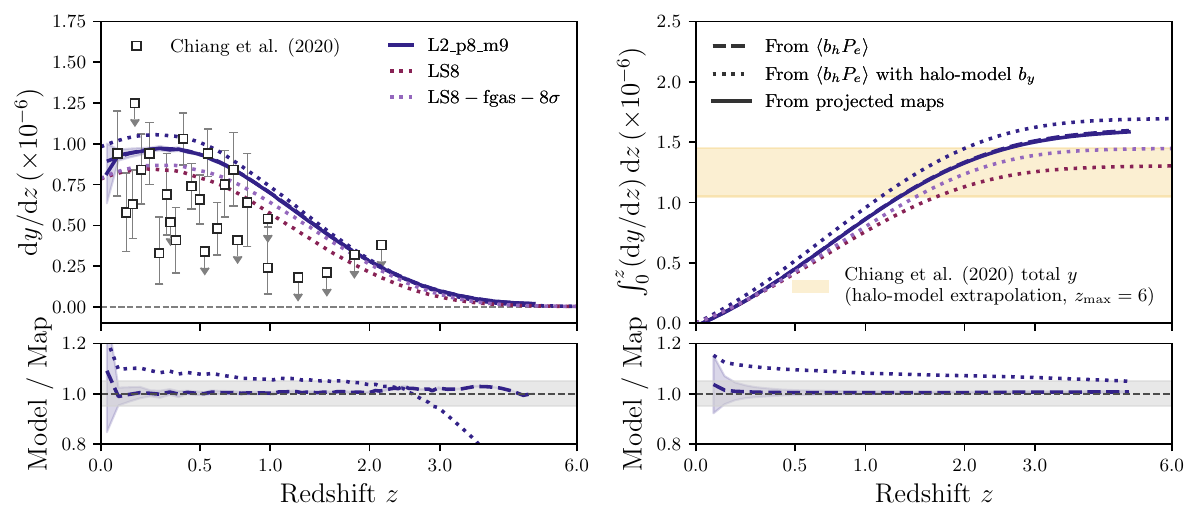}
  \caption{Comparison of the differential (\textbf{left}) and cumulative (\textbf{right}) thermal energy evolution obtained using three methods. The three methods are illustrated for the L2p8\_m9 fiducial run: the dashed lines show $\mathrm{d}y/\mathrm{d}z$ and $\left<y\right>(<z)$ derived from $\langle b_\mathrm{h} P_\mathrm{e} \rangle$ corrected with the simulation-measured $b_{y}$; the dotted lines show the same quantity corrected instead with the self-similar halo-model $b_{y}$ from \cref{eqn::b_y_halo_model}; and the solid lines are the corresponding values directly computed from the 2D projected lightcone maps. The shaded region around the solid line shows the spread obtained by averaging over 8 independent lightcones. In addition, the LS8 and LS8$-$fgas$-8\sigma$ models are shown using the halo-model $b_{y}$ correction (dotted lines); this method is directly comparable to the approach used by \citet{Chiang2020mzc} to infer $\mathrm{d}y/\mathrm{d}z$ and $\left<y\right>$ from their cross-correlation measurements, whose estimates are also shown for reference (data points in the left panel; orange shaded region in the right panel). The bottom sub-panels present the ratio of each method to the map-based estimate, with the grey band indicating agreement at the $\pm 5$ per cent level. The simulation-measured $b_{y}$ method and the maps agree to within a few per cent, while the halo-model $b_{y}$ method systematically overpredicts by $\approx5$ per cent owing to the lower bias predicted by the self-similar halo model (see \cref{by_comparison_plots}). A full summary of the sky-averaged $y$ monopole derived using all three methods is provided in \cref{tab:y_monopole_table}.}
  \label{dydz_3_methods}
\end{figure*}

\begin{table}
  \centering
  \caption{The tSZ $y$ monopole derived using three methods: (i) $\langle b_\mathrm{h} P_\mathrm{e} \rangle$ corrected with the simulation-measured $b_y$, cf.\ \cref{eqn::b_y_P_e_dy_dz_relation}, (ii) $\langle b_\mathrm{h} P_\mathrm{e} \rangle$ corrected with the self-similar halo-model $b_y$ from \cref{eqn::b_y_halo_model}, and (iii) direct calculations from the 2D projected lightcone maps. For all models, the predicted monopole is obtained by integrating up to $z = 3.0$.}
  \label{tab:y_monopole_table}
  \begin{tabular}{lccc}
    \hline
    Model & $\left<y\right>$ (sim.~$b_y$) & $\left<y\right>$ (HM~$b_y$) & $\left<y\right>$ (maps) \\
    \hline
    L1\_m9 (fiducial) & $1.49 \times10^{-6}$ & $1.56 \times10^{-6}$ & $1.48 \times10^{-6}$ \\
    \hline
    \multicolumn{4}{c}{Resolution and box size variations} \\
    L2p8\_m9 & $1.51 \times10^{-6}$ & $1.60 \times10^{-6}$ & $1.50 \times10^{-6}$ \\
    L1\_m10 & $1.51 \times10^{-6}$ & $1.59 \times10^{-6}$ & $1.50 \times10^{-6}$ \\
    L1\_m8 & $1.49 \times10^{-6}$ & $1.53 \times10^{-6}$ & $1.48 \times10^{-6}$ \\
    \hline
    \multicolumn{4}{c}{Cosmology variations} \\
    Planck & $1.55 \times10^{-6}$ & $1.62 \times10^{-6}$ & $1.54 \times10^{-6}$ \\
    PlanckNu0p24Var & $1.39 \times10^{-6}$ & $1.48 \times10^{-6}$ & $1.38 \times10^{-6}$ \\
    PlanckNu0p24Fix & $1.32 \times10^{-6}$ & $1.41 \times10^{-6}$ & $1.31 \times10^{-6}$ \\
    LS8 & $1.16 \times10^{-6}$ & $1.23 \times10^{-6}$ & $1.15 \times10^{-6}$ \\
    LS8$-$fgas$-8\sigma$ & $1.36 \times10^{-6}$ & $1.37 \times10^{-6}$ & $1.35 \times10^{-6}$ \\
    \hline
    \multicolumn{4}{c}{Feedback variations} \\
    fgas$+2\sigma$ & $1.45 \times10^{-6}$ & $1.53 \times10^{-6}$ & $1.44 \times10^{-6}$ \\
    fgas$-2\sigma$ & $1.54 \times10^{-6}$ & $1.59 \times10^{-6}$ & $1.53 \times10^{-6}$ \\
    fgas$-4\sigma$ & $1.59 \times10^{-6}$ & $1.63 \times10^{-6}$ & $1.58 \times10^{-6}$ \\
    fgas$-8\sigma$ & $1.74 \times10^{-6}$ & $1.72 \times10^{-6}$ & $1.73 \times10^{-6}$ \\
    NoCooling & $1.78 \times10^{-6}$ & $1.77 \times10^{-6}$ & $1.78 \times10^{-6}$ \\
    \hline
    \multicolumn{4}{c}{Other astrophysical variations} \\
    M*$-\sigma$ & $1.57 \times10^{-6}$ & $1.62 \times10^{-6}$ & $1.56 \times10^{-6}$ \\
    M*$-\sigma$\_fgas$-4\sigma$ & $1.69 \times10^{-6}$ & $1.70 \times10^{-6}$ & $1.68 \times10^{-6}$ \\
    Jet & $1.39 \times10^{-6}$ & $1.47 \times10^{-6}$ & $1.38 \times10^{-6}$ \\
    Jet\_fgas$-4\sigma$ & $1.84 \times10^{-6}$ & $1.79 \times10^{-6}$ & $1.83 \times10^{-6}$ \\
    \hline
  \end{tabular}
\end{table}

Using \cref{eqn::b_y_P_e_dy_dz_relation}, we derive the bias-corrected tomographic $\mathrm{d}y/\mathrm{d}z$ using the simulation-based measurements of $b_{y}$ and $\langle b_\mathrm{h} P_\mathrm{e} \rangle$ for each \flamingo\ model. To complement this, we also construct a halo-based estimate of the thermal history using the \flamingo\ halo lightcone catalogues. For each redshift shell, we sum the intrinsic integrated Compton $Y$ values of all haloes above a chosen mass threshold:
\begin{equation}
  \left.\frac{\mathrm{d}y}{\mathrm{d}z}\right|_{\rm halo}
  \simeq \frac{1}{4\pi\,\Delta z}\sum_{i\in \Delta z} \frac{Y_i(<R)}{D_\mathrm{A}^2(z_i)}\, ,
  \label{eq:dydz_halo_sum}
\end{equation}
where $Y_i(<R)$ is the intrinsic SO Compton $Y$ within aperture $R$ and $D_\mathrm{A}(z_i)$ the angular diameter distance. Gas particles heated directly by AGN within the last 15~Myr are excluded, as done for the $Y$--$M$ relation, though their inclusion has a negligible effect on the results.

\Cref{dydz_comparison_plots} shows the resulting tomographic $\mathrm{d}y/\mathrm{d}z$ for different model variations (solid curves), along with the corresponding halo-based reconstruction for haloes with $M_{500\mathrm{c}}\geq 10^{11}\,\Msol$ using the $r_{200\mathrm{c}}$ aperture (dashed curves). In general, we observe similar trends between the simulations in $\mathrm{d}y/\mathrm{d}z$ as previously observed in $\langle b_\mathrm{h} P_\mathrm{e} \rangle$.  For example, the fiducial model (L1\_m9) predicts a higher amplitude than the values reported by \citet{Chiang2020mzc}, who used the halo model to compute $b_y$ and thereby infer $\mathrm{d}y/\mathrm{d}z$ from their measurements of $\langle b_\mathrm{h} P_\mathrm{e} \rangle$. Because these estimates are model-dependent, they should be interpreted with caution rather than used to rule out specific simulation models. The LS8 model yields predictions that are closer to these data points, and stronger feedback tends to enhance $\mathrm{d}y/\mathrm{d}z$ rather than suppressing it.

Examining the dashed curves, which show the halo contribution to $\mathrm{d}y/\mathrm{d}z$, we find that slightly less than half of the signal comes from within $r_{200\mathrm{c}}$.  In Appendix \ref{sec:dydz_radius} we explore different radial aperture cuts and how the halo contribution to $\mathrm{d}y/\mathrm{d}z$ depends on baryon physics and cosmology.

In \cref{dydz_3_methods} we compare the differential $\mathrm{d}y/\mathrm{d}z$ (left panel) and the corresponding cumulative sky-averaged $y$ monopole (right panel) obtained using three approaches: (i) $\langle b_\mathrm{h} P_\mathrm{e} \rangle$ corrected with the simulation-measured $b_y$, (ii) $\langle b_\mathrm{h} P_\mathrm{e} \rangle$ corrected with the self-similar halo-model $b_y$ from \cref{eqn::b_y_halo_model} (with $\alpha_p = 0$), and (iii) direct calculations from the 2D projected lightcone maps. For the L2p8\_m9 run, which provides eight independent lightcone map outputs allowing us to estimate the impact of cosmic variance (shown as the shaded region), we show results from all three methods. We also include the LS8 and LS8$-$fgas$-8\sigma$ models using the halo-model $b_y$ correction (dotted lines), as this approach is directly comparable to the method employed by \citet{Chiang2020mzc} to derive their observational estimates.

The simulation-measured $b_y$ method and the maps yield consistent results, except for minor discrepancies at very low redshift driven by cosmic variance. The halo-model $b_y$ method, however, systematically overpredicts the monopole because the self-similar halo model predicts a lower bias than the actual measured values (see \cref{by_comparison_plots}), resulting in a $\approx5$ per cent higher inferred $y$ monopole \citep[see also][]{Chiang2020mzc}. For the cumulative $y$ monopole, all three methods approximately converge by $z\approx2$.  \Cref{tab:y_monopole_table} summarises the $y$ monopole obtained by integrating up to $z = 3.0$ for all \flamingo\ models using all three methods. The simulation-measured $b_y$ and map-based estimates are in excellent agreement, whereas the halo-model $b_y$ values are systematically higher, confirming that the choice of bias correction is a mild source of systematic uncertainty when inferring the $y$ monopole from cross-correlation measurements.

The $y$ monopole estimate obtained by \citet{Chiang2020mzc} is $1.22^{+0.23}_{-0.17} \times10^{-6}$, shown as the orange shaded region in the right panel of \cref{dydz_3_methods}. The fiducial run predicts a slightly higher value than this (model-dependent) tomographic measurement. Simulations with stronger feedback (lower target gas fractions), as well as the NoCooling model, all in the fiducial D3A cosmology, predict some of the largest $y$ monopoles (see \cref{tab:y_monopole_table}), driven either by enhanced thermal energy in feedback-ejected gas or by the absence of gas cooling and star formation, respectively.  In contrast, the LS8 model, and its stronger feedback variant LS8$-$fgas$-8\sigma$, yield the lowest values and are both in excellent agreement with the \citet{Chiang2020mzc} value, as can be seen directly in \cref{dydz_3_methods}.  As expected, these trends are consistent with the behaviour previously discussed in the comparisons of $\langle b_\mathrm{h} P_\mathrm{e} \rangle$  and $\mathrm{d}y/\mathrm{d}z$.

While estimates of the $y$ monopole can only be obtained in a model-dependent way from cross-correlations, a \textit{direct} constraint on $\left<y\right>$ can be placed using spectral distortion measurements of the CMB with spectrometers such as COBE-FIRAS. A recent re-analysis of the COBE-FIRAS data using an improved astrophysical foreground cleaning technique has reduced the upper limit by approximately a factor of three, from $<15\times10^{-6}$ \citep{COBE_FIRAS} to $5.2\times10^{-6}$ \citep{new_FIRAS_measure}.  This limit is still not sufficiently stringent to rule out any of the \flamingo\ models, but planned spectral distortion experiments (referenced in \cref{sec:intro}) promise to yield informative direct measurements in the near future.

\citet{new_FIRAS_measure} also compared the updated COBE-FIRAS upper limits with predictions from the CAMELS-SIMBA model variants \citep{CAMELS_ref1,CAMELS_ref2}, yielding predicted $\left<y\right>$ values in the range from $10^{-5}$ to $10^{-6}$ depending on the adopted strengths of AGN and supernova feedback. These authors found that increasing the AGN feedback strength generally raises $y$ monopole values (consistent with the present study), whereas increasing the supernova feedback strength lowers it. This supernova feedback dependency appears to be opposite to what we find when comparing the fiducial run with the M*$-\sigma$ run, though a direct comparison is not straightforward: in \flamingo, all four subgrid parameters are recalibrated simultaneously when shifting the target stellar mass function, so the M*$-\sigma$ variation does not isolate the effect of a single feedback channel.

\subsection{The imprint of feedback on tSZ scaling relations}
\label{ssec:scaling_relations}

Here we explore why stronger feedback enhances the signals in $\langle b_\mathrm{h} P_\mathrm{e} \rangle$ and $\mathrm{d}y/\mathrm{d}z$ at higher redshift, rather than suppressing them.

To illustrate this, we examine the $Y$--$M$ relation derived from the simulated halo catalogues (see also \citealt{Kugel2024}). The Compton $y$ value for each halo is calculated by summing the contributions from gas particles within the SO-defined sphere (see \cref{eqn::FMG_y} for the definition of the Compton $y$ parameter for an individual gas particle). Contributions from gas particles that have recently been directly heated by AGN feedback are excluded.

\Cref{Y_M_fgas_test_0_1} and \cref{Y_M_fgas_test_1.5_3} show the median $Y$--$M$ relation calculated within two apertures, $r_{500\mathrm{c}}$ and $5r_{500\mathrm{c}}$, with the latter capturing partial contributions from the two-halo regime. To characterise the redshift evolution of this relation, we analyse six halo catalogues at $z=$ 0.1, 0.5, 1.0, 1.5, 2.0, and 3.0. Here we focus on the results obtained from the gas fraction variations (fgas variants). For comparison, we also show the self-similar prediction, as well as the NoCooling model with its best-fitting $Y$--$M$ power-law index.

It is evident that the best-fitting power-law index for the NoCooling model deviates significantly from the self-similar expectation, particularly for the $5r_{500\mathrm{c}}$ case at higher redshift. As discussed in Section~\ref{ssec:integrated_tsz}, this explains the deviation in the redshift evolution of $b_{y}$ at $z\gtrsim2.0$ with respect to the self-similar halo model. For the NoCooling model, we find that the deviation from the self-similar expectation of $Y \propto M^{5/3}$ is driven by a deviation in the (mass-weighted) temperature--halo mass relation from the virial theorem expectation of $T \propto M^{2/3}$, which in turn is likely due to deviations from virial equilibrium at early times and large radii.

Examining the $Y_{r_{500\mathrm{c}}}$--$M$ relation across different feedback models, we find that within the one-halo regime, feedback efficiently expels hot gas from haloes, leading to a suppression of the integrated $Y$ signal across a broad range of halo masses. When including gas from the two-halo regime, as per the $Y_{5r_{500\mathrm{c}}}$--$M$ relation, a different trend emerges at high redshift: stronger feedback increases the integrated $Y$ within $5r_{500\mathrm{c}}$ compared to weaker feedback models. This indicates that feedback effectively heats the gas and redistributes thermal energy into the surrounding environment, enhancing the contribution from the two-halo regime. In principle, this prediction can be tested observationally, for example by stacking samples of high-redshift galaxies to measure their mean tSZ profiles.

\section{Summary and conclusions}
\label{sec:conclusions}

We have used the \flamingo\ suite of cosmological hydrodynamical simulations to study the bias-weighted mean electron pressure $\langle b_\mathrm{h} P_\mathrm{e} \rangle$, the thermal history $\mathrm{d}y/\mathrm{d}z$, and the effective $y$-weighted halo bias $b_y$ across 18 simulation variants spanning different cosmologies and baryonic feedback implementations. Cross-correlating tSZ intensity maps with galaxy or quasar surveys yields a direct measurement of $\langle b_\mathrm{h} P_\mathrm{e} \rangle$ as a function of redshift, where $b_\mathrm{h}$ is the linear halo bias, quantifying how the massive haloes that dominate the thermal pressure cluster relative to the underlying matter field, and $P_\mathrm{e}$ is the mean electron pressure. With a model for the $y$-weighted halo bias $b_y$, one can further derive the thermal history $\mathrm{d}y/\mathrm{d}z$, which characterises the redshift evolution of the Compton $y$ parameter. Our main findings are:

\begin{itemize}

  \item $\langle b_\mathrm{h} P_\mathrm{e} \rangle$ is well converged with resolution and box size in \flamingo, with differences below 5 per cent between the intermediate (m9) and high (m8) resolution runs (top left panel of \cref{fig:Pe_data}).

  \item $\langle b_\mathrm{h} P_\mathrm{e} \rangle$ depends steeply on cosmology, scaling as $S_8^{\epsilon(z)}$ with an effective exponent $\epsilon(z) \approx 3$ over the redshift range $0.1 \leq z \leq 1$ (see \cref{fig:s8}). The Planck cosmology ($S_8 = 0.833$) systematically overpredicts the observational data, while the LS8 cosmology ($S_8 = 0.766$) provides the best fit (top right panel of \cref{fig:Pe_data}).

  \item Baryonic feedback variations are subdominant at low redshift but become increasingly important at higher redshifts (bottom panels of \cref{fig:Pe_data}). Stronger feedback \emph{increases} $\langle b_\mathrm{h} P_\mathrm{e} \rangle$ by redistributing thermal energy to larger scales (see \cref{Y_M_fgas_test_0_1} and \cref{Y_M_fgas_test_1.5_3}) --- opposite to the effect of feedback in cosmic shear analyses.

  \item We find that the \flamingo\ model with a low $S_8$ cosmology and strong feedback (i.e., LS8$-$fgas$-8\sigma$) provides the best fit to the observational measurements of $\langle b_\mathrm{h} P_\mathrm{e} \rangle$, consistent with recent studies examining the tSZ power spectrum \citep{Ian_kSZ_feedback_FLAMINGO,SPT2026}, tSZ--cosmic shear cross-spectrum \citep{McCarthy_2023,Ian_kSZ_feedback_FLAMINGO}, and kSZ effect stacking measurements of massive galaxies \citep{Ian_kSZ_feedback_FLAMINGO,Siegel2026}.

  \item We introduce a compact phenomenological parameterisation of $\langle b_\mathrm{h} P_\mathrm{e} \rangle$ as a joint function of $S_8$ and the baryon fraction of group-mass haloes (\cref{eq:bhpe_s8_fb_ansatz}), which reproduces all \flamingo\ variants (excluding NoCooling) over $0.05 \leq z \leq 1$ with a mean error of $1.7 \pm 1.3$ per cent (worst case $\approx 7$ per cent; \cref{fig:test_fit}). An MCMC inference exercise using this model, marginalising over the model-parameter uncertainties, yields $S_8 = 0.72^{+0.03}_{-0.03}$ and a derived baryon fraction at $z=0.1$ of $f_\mathrm{b}(10^{13}\,\Msol, z{=}0.1)/(\Omega_\mathrm{b}/\Omega_\mathrm{m}) = 0.10^{+0.09}_{-0.05}$ (\cref{fig:compare_posteriors}). Splitting the data by redshift reveals that the low- and high-redshift regimes have complementary degeneracy directions in the $S_8$--$f_\mathrm{b}$ plane: the $z\leq0.3$ data constrain $S_8$ nearly independently of feedback, while the $z>0.3$ data are sensitive to both parameters but with a strong degeneracy between them. A complementary two-stage analysis that exploits this separation, first constraining $S_8$ from the low-redshift data and then fitting for the baryon fraction at high redshift, yields consistent results with a slightly higher $S_8 = 0.75 \pm 0.01$ and $f_\mathrm{b}(10^{13}\,\Msol,\,z{=}0.1)/(\Omega_\mathrm{b}/\Omega_\mathrm{m}) = 0.15^{+0.04}_{-0.03}$. These best-fitting values should, however, be interpreted with some caution: the preferred all-redshift solution lies slightly beyond the simulation grid used to calibrate the model and the absolute $\chi^2$ values are affected by data systematics that no smooth physical model can fully account for. The most robust conclusion is therefore that lower-$S_8$ models are preferred, with the lowest-$S_8$ simulation variants providing the best relative match to the measurements.  These results are broadly consistent with other recent tSZ-based analyses examining the tSZ power spectrum \citep{Bolliet2018,McCarthy_2023,Ian_kSZ_feedback_FLAMINGO,SPT2026} and the tSZ--cosmic shear cross-spectrum \citep{Troster2022,Ian_kSZ_feedback_FLAMINGO} but lie below that inferred from the primary CMB \citet{Planck2020cosmopars} and recent cosmic shear measurements (e.g., \citealt{Wright2025}).  See \cref{fig:s8_comparison}.

  \item The $y$-weighted halo bias increases with redshift as the tSZ signal shifts towards rarer, more massive systems (\cref{by_comparison_plots}).  Similar to the bias-weighted pressure, the $y$-weighted halo bias depends on both cosmology and baryon physics.  Consistent with the findings of \citet{Young2021}, we find that the hydrodynamical simulations deviate from the halo model prediction for $b_y$ at high redshift ($z\ga2$), which we attribute to deviations of the integrated $Y$--$M$ relation from self-similarity at high redshifts and large apertures (\cref{by_comparison_plots}).

  \item  The thermal history $\mathrm{d}y/\mathrm{d}z$ peaks at $z \approx 0.5$--1 and, similar to the bias-weighted pressure, depends on cosmology and baryon physics.  We show that the integrated $y$ monopole can be accurately inferred from measurements of the bias-weighted pressure together with an accurate model of $b_y$ (\cref{dydz_3_methods}).  Measurements of the $y$ monopole using this method by \citet{Chiang2020mzc} are in mild tension with \flamingo\ models in a Planck cosmology but in excellent agreement with models in a low $S_8$ cosmology.
    We also show that only $\approx 45$ per cent of $\mathrm{d}y/\mathrm{d}z$ signal originates within $r_{200\mathrm{c}}$ of haloes (\cref{fig:dydz_ratio_fid} and \cref{fig:dydz_ratio_models}). This fraction is controlled primarily by feedback (ranging from 29 to 51 per cent across models) rather than cosmology (43--46 per cent). A substantial fraction of the tSZ signal therefore arises from gas in halo outskirts and the diffuse intergalactic medium.

\end{itemize}

The comparison to existing measurements of the bias-weighted pressure and thermal history in this study assumed that the measurements contain no significant systematic uncertainties due to the presence of foreground contaminants such as the clustered infrared background (CIB) and radio point sources.  While techniques for minimising the effects of these contaminants are rapidly improving (e.g., \citealt{La_Posta_26}), a full forward modelling approach using foreground maps produced from the same hydrodynamical simulations that predict the thermal history (e.g., \citealt{Yang2026}) would be an interesting and important next step to take to establish the robustness of the measurements and the physical conclusions drawn from them.

Leaving aside these observational challenges, our results demonstrate that thermal pressure statistics provide a powerful and complementary probe of both the amplitude of matter fluctuations and baryonic feedback. The steep cosmology scaling, combined with the mild effects of feedback at low redshifts, make $\langle b_\mathrm{h} P_\mathrm{e} \rangle$ an attractive observable for constraining cosmological parameters that are sensitive to the growth of large-scale structure, including $S_8$. At higher redshifts, the growing sensitivity to feedback opens a window for constraining baryonic physics independently of probes such as cosmic shear.

The parameterisation developed here can be applied to observed cross-correlation measurements from current and forthcoming surveys, including DESI, Euclid, LSST, and Simons Observatory. Improved measurements of $\langle b_\mathrm{h} P_\mathrm{e} \rangle$ at $z > 1$ will provide stringent new tests of galaxy formation models and potentially sharpen constraints on cosmological parameters from an entirely independent channel.

\section*{Acknowledgements}
The authors thank David Alonso, Raul E. Angulo, Adrien La Posta, and Matteo Zennaro for helpful discussions. This work was supported by the Science and Technology Facilities Council (grant number ST/Y002733/1).
This project has received funding from the European Research Council (ERC) under the European Union's Horizon 2020 research and innovation programme (grant agreement No 769130). This work was supported by the Science and Technology Facilities Council [ST/P000541/1]. This work used the DiRAC@Durham facility managed by the Institute for Computational Cosmology on behalf of the STFC DiRAC HPC Facility (www.dirac.ac.uk). The equipment was funded by BEIS capital funding via STFC capital grants ST/K00042X/1, ST/P002293/1 and ST/R002371/1, Durham University and STFC operations grant ST/R000832/1. DiRAC is part of the National e-Infrastructure. AU acknowledges the support of the National Natural Science Foundation of China (NSFC) under grant No. 12473005, and the Yunnan Key Laboratory of Survey Science, Project No. 202449CE340002.

\section*{Data availability}
A detailed description of the relevant \flamingo\ data products and access arrangements is given by \citet{Helly2026}. The data underlying this article may be shared on reasonable request to the corresponding author.




\bibliographystyle{mnras}
\bibliography{newbib} 




\appendix
\crefalias{section}{appendix}

\section{Validation of parametric model of bias-weighted pressure}
\label{sec:validation}

\begin{figure*}
  \centering
  \includegraphics[width=0.98\textwidth]{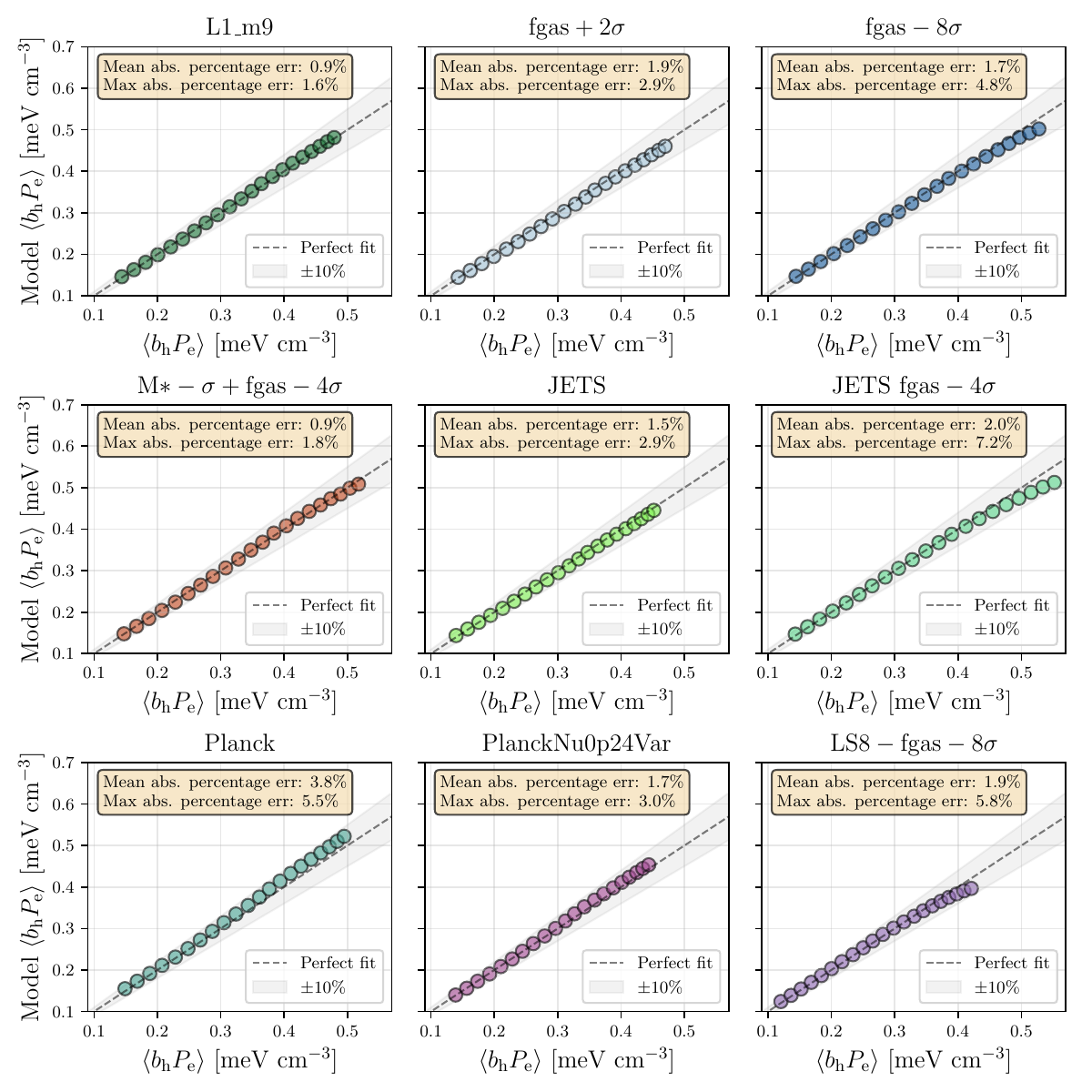}
  \caption{Validation of the joint cosmology and baryon fraction parameterisation in \cref{eq:bhpe_s8_fb_ansatz} against direct power-spectrum measurements for a representative subset of nine \flamingo\ variants selected to span the full range of feedback strengths and cosmologies, including the cases with the largest residuals: the fiducial model (L1\_m9), the weak feedback variant (fgas$+2\sigma$), the strongest thermal feedback variant (fgas$-8\sigma$), the combined stellar mass and gas fraction variation (M*$-\sigma$+fgas$-4\sigma$), the jet-mode AGN feedback model (Jets), the strong jet-mode variant (Jets fgas$-4\sigma$), the Planck cosmology, the Planck cosmology with massive neutrinos (PlanckNu0p24Var), and the LS8 cosmology with the strongest feedback (LS8$-$fgas$-8\sigma$). In each panel, points show $\langle b_\mathrm{h} P_\mathrm{e} \rangle$ evaluated directly from the matter--electron pressure cross-power spectrum ($x$-axis) against the corresponding prediction of \cref{eq:bhpe_s8_fb_ansatz} with the coefficients in \cref{tab:s8_fb_fit_params} ($y$-axis), for all redshift snapshots in $0.05 \leq z \leq 1$. The dashed line indicates the 1:1 relation and the shaded band marks $\pm 10$ per cent. Evaluated over all 14 variants (see text), the mean absolute relative error is $1.7 \pm 1.3$ per cent (median 1.4 per cent; worst case $\approx 7$ per cent), with the residuals growing mildly from $\approx 1.4$ per cent at $z \leq 0.3$ to $\approx 2.1$ per cent at $z \approx 0.7$--$1$.}
  \label{fig:test_fit}
\end{figure*}

Here we test our simple parameterisation in \cref{eq:bhpe_s8_fb_ansatz} for the effects of cosmology (particularly $S_8$) and baryon physics (captured with the group baryon mass fraction) on the bias-weighted pressure. \Cref{fig:test_fit} shows the comparison for a representative subset of nine variants, selected to span the full range of feedback strengths and cosmologies and to include the cases with the largest residuals. The error statistics quoted below are computed over all 14 \flamingo\ simulation variants (excluding only box size, resolution, and adiabatic variations). Across all variants and redshifts, the mean absolute relative error is $1.7 \pm 1.3$ per cent (median 1.4 per cent), with a worst-case deviation of $\approx 7$ per cent. The residuals show a slight increase in scatter with redshift (i.e., mild heteroscedasticity): the mean error is lowest at low redshift ($\approx 1.4$ per cent for $z \leq 0.3$), $\approx 1.5$ per cent at intermediate redshift ($0.33 < z \leq 0.66$), and rises to $\approx 2.1$ per cent at $0.66 < z \leq 1$, where the sensitivity to both cosmological and baryonic parameters is greatest.

\section{Dependence of thermal history on halo radial extent}
\label{sec:dydz_radius}
Here we decompose the thermal history $\mathrm{d}y/\mathrm{d}z$ by halo radial extent using the halo lightcone-based reconstruction defined in \cref{eq:dydz_halo_sum}, in which the intrinsic Compton $Y$ values of individual haloes from the \flamingo\ halo lightcone catalogues are summed within different radial apertures.

\begin{figure}
  \centering
  \includegraphics[width=0.48\textwidth]{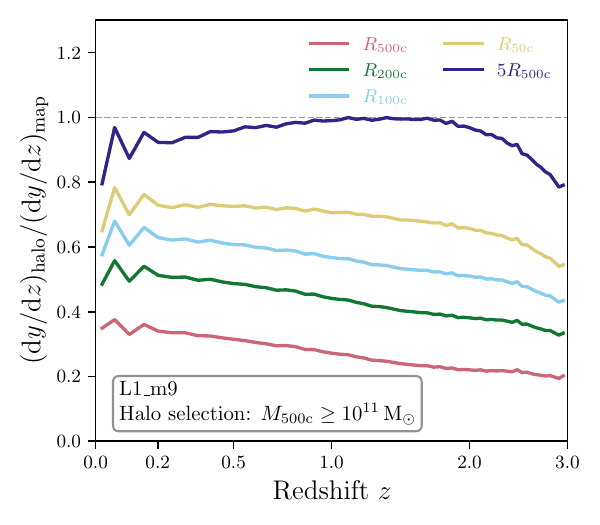}
  \caption{Halo-to-map ratio of the tSZ lightcone signal in the fiducial \flamingo\ model L1\_m9. The quantity shown is $(\mathrm{d}y/\mathrm{d}z)_\mathrm{halo} /(\mathrm{d}y/\mathrm{d}z)_\mathrm{map}$ for haloes with $M_{500\mathrm{c}}\geq 10^{11}\,\Msol$, comparing five apertures used to define the halo signal: $r_{500\mathrm{c}}$, $r_{200\mathrm{c}}$, $r_{100\mathrm{c}}$, $r_{50\mathrm{c}}$, and $5r_{500\mathrm{c}}$. Enlarging the aperture substantially increases the fraction of the total tSZ signal arising within the halo apertures at all redshifts, showing that halo outskirts contribute an important part of the tSZ budget. However, the ratio remains below unity even for $5r_{500\mathrm{c}}$, indicating that not all of the tSZ background is accounted for by very extended halo apertures. The corresponding redshift-integrated ratios are 0.28, 0.45, 0.57, 0.70, and 0.96 for $r_{500\mathrm{c}}$, $r_{200\mathrm{c}}$, $r_{100\mathrm{c}}$, $r_{50\mathrm{c}}$, and $5r_{500\mathrm{c}}$, respectively.}
  \label{fig:dydz_ratio_fid}
\end{figure}

\Cref{fig:dydz_ratio_fid} quantifies the dependence on halo radial extent by showing the ratio $(\mathrm{d}y/\mathrm{d}z)_\mathrm{halo}/(\mathrm{d}y/\mathrm{d}z)_\mathrm{map}$ for five aperture choices. The redshift-integrated halo contribution increases from 28 per cent within $r_{500\mathrm{c}}$ to 45 per cent within $r_{200\mathrm{c}}$, 57 per cent within $r_{100\mathrm{c}}$, 70 per cent within $r_{50\mathrm{c}}$, and 96 per cent within $5r_{500\mathrm{c}}$. Thus, the majority of the thermal history signal in the full lightcone can be accounted for by the tSZ signal within $5r_{500\mathrm{c}}$ of haloes, though a residual $\approx 4$ per cent remains outside even the $5r_{500\mathrm{c}}$ aperture.

\begin{figure*}
  \centering
  \includegraphics[width=0.98\textwidth]{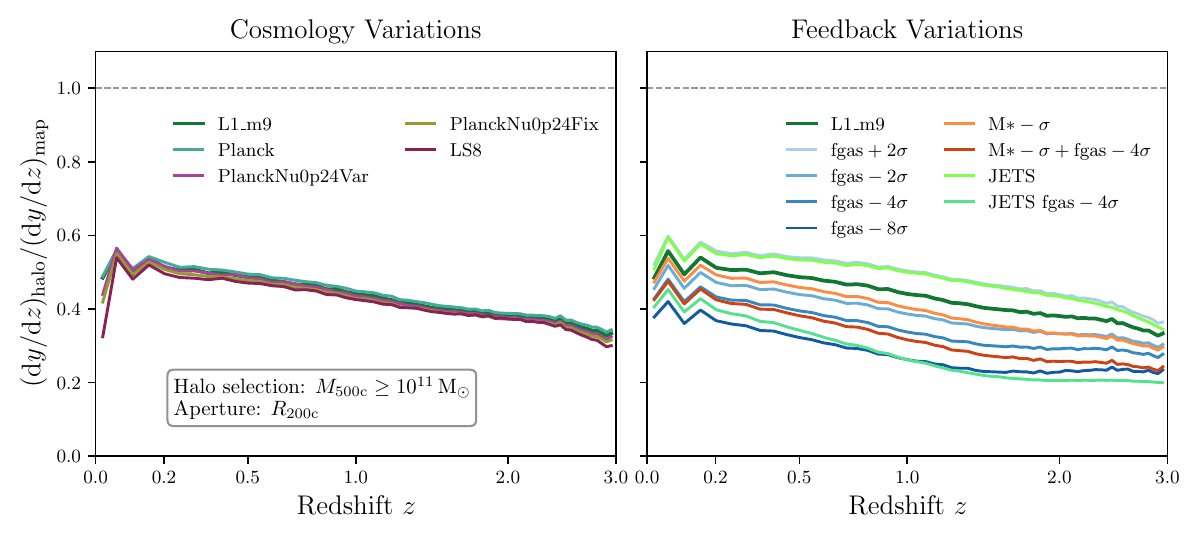}
  \caption{Model dependence of the halo-to-map ratio of the tSZ lightcone signal at fixed halo definition. The quantity shown is $(\mathrm{d}y/\mathrm{d}z)_\mathrm{halo} /(\mathrm{d}y/\mathrm{d}z)_\mathrm{map}$ for haloes with $M_{500\mathrm{c}}\geq 10^{11}\,\Msol$, using the aperture $r_{200\mathrm{c}}$. \textbf{Left}: Cosmology variations. \textbf{Right}: Baryonic-physics variations. The fraction of the total tSZ signal arising within the halo apertures changes only weakly across the cosmology variants, but varies strongly across the feedback models. Stronger feedback systematically lowers the contribution from within halo apertures, consistent with hot gas being displaced to larger radii and lower-density environments. The redshift-integrated ratio ranges from 0.434 to 0.460 across the cosmology variants, but from 0.287 to 0.506 across the feedback variations.}
  \label{fig:dydz_ratio_models}
\end{figure*}

The comparison across variants (\cref{fig:dydz_ratio_models}) confirms that this ratio is controlled primarily by feedback rather than cosmology. At fixed $r_{200\mathrm{c}}$, the integrated ratio varies only from 43.4 to 46.0 per cent across cosmology variants, but from 28.6 to 50.6 per cent across feedback variations. Stronger feedback displaces hot gas to larger radii, lowering the within-halo fraction, while cosmology changes primarily rescales the overall amplitude without strongly altering the halo-to-total partition.

\section{Boxsize convergence test for the bias-weighted mean electron
pressure}\label{appen:Boxsize_convergence_test}

\begin{figure*}
  \centering
  \includegraphics[width=0.98\linewidth]{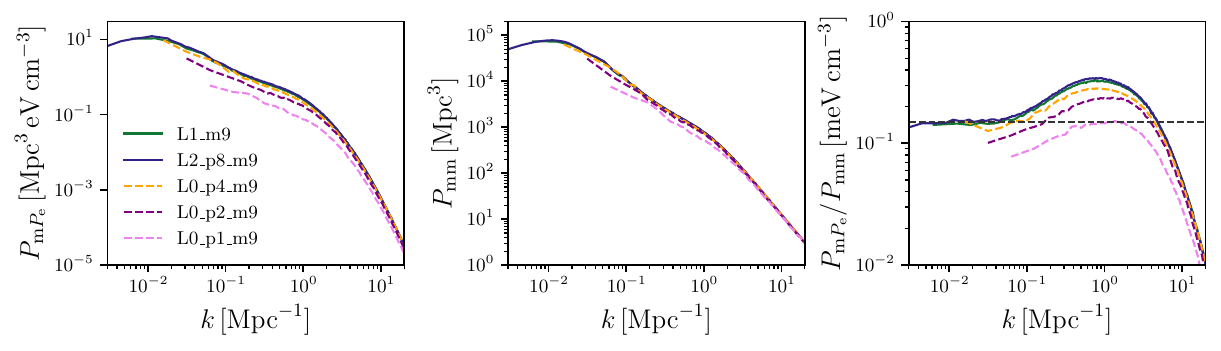}
  \caption{Comparison of three-dimensional power spectra at $z = 0$ from FLAMINGO fiducial run with box sizes of 100 Mpc (denoted as L0p1\_m9 in violet dashed curve), 200 Mpc (denoted as L0p2\_m9 in purple dashed curve), 400 Mpc (denoted as L0p4\_m9 in orange dashed curve), 1 Gpc, and 2.8 Gpc (all at fixed resolution). \textbf{Left}: matter--electron pressure cross-power spectrum $P_{\rm  mP_{\mathrm{e}}}(k,z=0)$. \textbf{Middle}: matter power spectrum $P_{\rm mm}(k,z=0)$. \textbf{Right}: ratio between two 3D power spectra, with large-scale limit used for computing the bias-weighted mean pressure $\langle b_\mathrm{h} P_\mathrm{e} \rangle$. The horizontal dashed line indicates the mean large-scale plateau value inferred from the 1 Gpc and 2.8 Gpc simulations. Convergence is achieved for $L>400~\rm Mpc$, while smaller boxes systematically underestimate the large-scale amplitude and fail to recover a stable constant limit in the power spectra ratio.}
  \label{convergence_test_PS_bPe}
\end{figure*}

To assess the robustness of our large-scale $\langle b_\mathrm{h} P_\mathrm{e} \rangle$ estimator, we compare the 3D matter--electron pressure cross-power spectrum, the matter auto-power spectrum, and their ratio at $z = 0$ across FLAMINGO fiducial simulations with different box sizes (100, 200, 400 Mpc, 1 Gpc, and 2.8 Gpc), all at the same  resolution.

The left panel of \cref{convergence_test_PS_bPe} shows the 3D matter–pressure power spectrum,
$P_{\rm m P_{\mathrm{e}}}(k,z=0)$. The 1 Gpc and 2.8 Gpc boxes are in excellent agreement over a wide range of $k$. The 400 Mpc box case also reproduces the overall shape and amplitude reasonably well, though with mild fluctuations toward the largest scales due to limited volume. In contrast, the 200 Mpc and especially the 100 Mpc box cases predict a much lower amplitude across a broad range of scales, which indicates finite-volume effects as well as the absence of massive clusters and galaxy groups in smaller simulation boxes.

The middle panel shows the matter power spectrum, $P_{\rm mm}(k,z=0)$. Here, the agreement is much better across box sizes, with only the 100 Mpc box showing a slight suppression at the largest scales. The right panel presents the ratio $P_{\rm mP_{\mathrm{e}}}$/$P_{\rm mm}$, which corresponds to the bias-weighted mean pressure $\langle b_\mathrm{h} P_\mathrm{e} \rangle$ in the large-scale limit. The 1 Gpc and 2.8 Gpc results are fully consistent and reach a constant plateau at small $k$. The 400 Mpc box yields a broadly similar trend, though the large-scale end is noisier compared to larger box sizes. The 200 Mpc and 100 Mpc boxes produce significantly lower amplitudes and do not exhibit clear convergence toward a constant large-scale plateau. For these smaller volumes, the large-scale limit is not reliably established. From the comparison of $P_{\rm m P_{\mathrm{e}}}$ and $P_{\rm mm}$, this arises primarily from the matter–pressure cross-correlation in smaller box sizes rather than from the total matter distribution itself.

These results demonstrate that the large-scale-limit method for estimating $\langle b_\mathrm{h} P_\mathrm{e} \rangle$ is intrinsically volume-dependent. Reliable convergence requires simulation boxes with side length $L>400~\rm Mpc$. For smaller volumes (e.g. 100–200 Mpc as examined here), the absence of long-wavelength modes leads to a biased and suppressed estimate of the large-scale amplitude, which makes a direct determination of the constant $\langle b_\mathrm{h} P_\mathrm{e} \rangle$ from 3D power spectra nearly impossible.

We therefore emphasise that caution is required when applying this approach to small-volume simulations, particularly in studies aiming to perform thermal energy analyses using the large-scale-limit method. Without sufficient volume, the inferred large-scale bias-weighted pressure can be systematically underestimated.


\bsp  
\label{lastpage}
\end{document}